\documentclass{article}

\usepackage{PRIMEarxiv}

\usepackage[utf8]{inputenc} 
\usepackage[T1]{fontenc}    
\usepackage{hyperref}       
\usepackage{url}            
\usepackage{booktabs}       
\usepackage{amsfonts}       
\usepackage{nicefrac}       
\usepackage{microtype}      
\usepackage{lipsum}
\usepackage{fancyhdr}       
\usepackage{graphicx}       
\usepackage{amssymb}
\usepackage{amsmath}
\graphicspath{{media/}}     

\title{Honest Score Client Selection Scheme: Preventing Federated Learning Label Flipping Attacks in Non-IID Scenarios
}

\author{
  Yanli Li \textsuperscript{1}, Huaming Chen\textsuperscript{2}, Wei Bao\textsuperscript{3}, Dong Yuan\textsuperscript{5} \\
  School of Electrical and Information Engineering\\
  The University of Sydney \\
  Sydney, NSW\\
  \texttt{\{yanli.li, huaming.chen, wei.bao, dong.yuan\}@sydney.edu.au} \\
   \And
  Zhengmeng Xu\textsuperscript{4}\ \\
  Postdoctoral Research Center \\
  Industrial and Commercial Bank of China \\
  Beijing\\
  \texttt{sddxxzm@126.com} \\
}

\begin{document}
\maketitle

\begin{abstract}
Federated Learning (FL) is a promising technology that enables multiple actors to build a joint model without sharing their raw data. The distributed nature makes FL vulnerable to various poisoning attacks, including model poisoning attacks and data poisoning attacks. Today, many byzantine-resilient FL methods have been introduced to mitigate the model poisoning attack, while the effectiveness when defending against data poisoning attacks still remains unclear. In this paper, we focus on the most representative data poisoning attack - “label flipping attack” and monitor its effectiveness when attacking the existing FL methods. The results show that the existing FL methods perform similarly in Independent and identically distributed (IID) settings but fail to maintain the model robustness in Non-IID settings. To mitigate the weaknesses of existing FL methods in Non-IID scenarios, we introduce the Honest Score Client Selection (short as HSCS) scheme and the corresponding HSCSFL framework. In the HSCSFL, The server collects a clean dataset for evaluation. Under each iteration, the server normally collects the gradients from clients and then perform HSCS to select aggregation candidates. Specially, the server first evaluates the performance of each class of the global model and generates the corresponding risk vector to indicate which class could be potentially attacked. Similarly, the server evaluates the client's model and records the performance of each class as the accuracy vector. The dot product of each client's accuracy vector and global risk vector is generated as the client's host score; only the top $p\%$ host score clients are included in the following aggregation. Once candidates have been selected by HSCS, server aggregates the gradients and uses the outcome to update the global model. The experiments are conducted on MNIST and Fashion MNIST with different Non-IID degrees and “label flipping attack” strategies. The empirical results show our HSCSFL effectively enhances the FL robustness and defends against the “label flipping attack.”
\end{abstract}

\keywords{Federated Learning \and Data Poisoning Attack \and Client Selection}

\section{Introduction}
Federated Learning (FL) \cite{mcmahan2017communication} is a distributed machine learning (ML) paradigm that enables many clients to jointly train a model under a service provider (i.e., server) without sharing the raw data. Today, FL has been wildly deployed in different domains and corporations, including the smartphone's intelligence input “Gboard” of Google \cite{yang2018applied}, the virtual assistant “Siri” of Apple \cite{hao2020apple}, “Alexa” of Amazon \cite{chen2022actperfl}, and drug discovery in the pharmaceutical industries \cite{chen2020fl}. The standard FL \cite{mcmahan2017communication} iteration includes 3 stages: (1) the server broadcasts the current global model to all participants. (2) the participants locally train the model through the privacy training data and send the local model updates back to the server. (3) the server averages the gradients\footnote{In this work, we combined use “model update” and “gradient” with the same meaning.} received to generate the global model gradient and consequently update the global model to start the next learning iteration. To speed up the FL's training while keeping a high learning performance, some recent FL works \cite{li2022pyramidfl,zhang2022multi,huba2022papaya, luo2021cost} consider introducing a “client selection” stage before each iteration. Under such FL frameworks, only the clients that satisfy the selection rule will participate in the aggregation; others are excluded from the current learning iteration.    

Although the Federated Learning system maintains the participants' privacy by enabling clients locally train their model, it introduces a new attack surface for the adversary and makes FL vulnerable to poisoning attacks \cite{zhang2022challenges,gosselin2022privacy, mothukuri2021survey, kairouz2021advances}. In particular, the adversary can poison the global model and degrade its performance by uploading crafted gradients. Examples of poisoning attacks include “model poisoning attack” \cite{cao2022mpaf, fang2020local, baruch2019little,xie2020fall} and “data poisoning attack” \cite{lewis2023attacks, yerlikaya2022data, tolpegin2020data} which directly manipulates the report and crafts the training data to generate the poisoning gradient, respectively. As the most representative data poisoning attack, the label flipping attack has been introduced by \cite{lewis2023attacks, tolpegin2020data}. In “label flipping attack”, the adversary first agrees on the attack strategy, which includes the attack class(es), target class, attack timing, etc. Then, the adversary flips the attack class of training data to the target class and uses the crafted training data to train the model, the poisoned gradient is consequently generated and submitted to the server.    

To defend against the poisoning attacks and enhance the robustness of the FL system, Byzantine resilient federated learning has been introduced by recent researchers \cite{li2022enhancing,cao2020fltrust,yin2018byzantine,blanchard2017machine}. By leveraging the statistics evaluation, these works decrease the weight of suspicious client in aggregation or selecting the most honest client as the aggregation outcome. For instance, Krum \cite{blanchard2017machine} calculates the sum Euclidean distance for a client's gradients received and selects the gradient achieving the shortest distance as the aggregation outcome in the current iteration. However, we note that most of these FL methods are designed for model poisoning attacks. As data poisoning attacks introduce smaller perturbations, the performance when defending against data poisoning attack, especially label flipping attack \cite{lewis2023attacks, tolpegin2020data}, still remains unclear. Furthermore, the learning scenario of real-world FL applications is usually Non-Independent and Identically Distributed (Non-IID) \cite{kairouz2021advances, li2022enhancing}. The clients' training data and corresponding training models are usually heterogeneous due to the different user habits, locations, time windows, etc. For instance, people’s shopping patterns are different due to the personal habit and design trends, the mammal’s distributions are different due to the geographic location, the images of the parked cars sometimes are snow-covered due to the seasonal effects \cite{mothukuri2021survey}. The Non-IID nature leads the huge difference between the models of various participants, bringing difficulty for the existing FL methods to identify the adversary and defend against poisoning attacks.

In this paper, we first evaluate the robustness of the existing Byzantine resilient FL methods \cite{cao2020fltrust,yin2018byzantine,blanchard2017machine} when defending against “label flipping attack” \cite{lewis2023attacks, tolpegin2020data} in both IID and Non-IID settings. The experimental results show that the existing FL methods can maintain the robustness in IID scenarios but fails to defend against such attacks \cite{lewis2023attacks, tolpegin2020data} under Non-IID settings. To mitigate the research gap, we propose the Honest Score Client Selection scheme (short as HSCS) and corresponding Honest Score Client Selection FL (short as HSCSFL). In HSCSFL, the server collects a clean evaluation dataset which evenly include data of each class. During each learning iteration, the server first evaluates and records each class's performance of the global model, the class achieving low accuracy is consequently assigned a high risk value. We call the recording of global model performance and the corresponding risk values as performance vector ($PerV$) and risk vector ($RisV$) respectively. Then, HSCSFL evaluates the class performance of each clients' model and generates the accuracy vector ($AccV$) for each client, Finally, the server calculates the dot product of the risk vector and the accuracy vector for each client as its honest score ($HS$), only the client with top $p\%$ honest score are included in the following averaging aggregation. We evaluate our HSCSFL in MNIST and FMNIST with different Non-IID degrees and different label flipping strategies, the experimental result shows our HSCSFL can maintain the robustness when defense against “label flipping attack.” \cite{lewis2023attacks, tolpegin2020data} 

Our contributions are summarized as follows: 
\begin{itemize}
\item We comprehensively evaluate the performance of existing FL methods when defending against “label flipping attack” in both IID and Non-IID settings. Our results show these methods witness a degrading performance in Non-IID settings. 
\item We propose the Honest Score Client Selection scheme, which can effectively select the most honest clients in the current learning iteration.
\item We propose the Honest Score Client Selection Federated Learning; our result shows HSCSFL can effectively defense against “label flipping attack.” 
\end{itemize}

\section{Related Work}
\subsection{Federated Learning Applications in Real-world}
As a collaborative training framework for machine learning that safeguards privacy, FL has garnered widespread attention from the industry. Today, numerous real-world FL applications have been proposed and deployed. 
These applications can generally be divided into three categories, including applications for smart devices \cite{yang2018applied, sozinov2018human, ramaswamy2019federated}, industrial engineering \cite{hu2018federated, han2019visual} and healthcare domains \cite{kim2017federated, huang2019patient}.

First, with Google's successful deployment of the user input prediction keyboard (Gboard) based on federated learning, applications of federated learning in smart devices have become a hot exploration concept.This category of applications includes not only text prediction of user input \cite{yang2018applied} but also the prediction of emojis \cite{ramaswamy2019federated} and the behavior of device users \cite{sozinov2018human}. Second, given the achievement of FL in privacy preservation, an increasing number of FL based industrial engineering applications are being developed and deployed, especially in data-sensitive fields within industrial engineering. For instance, the study \cite{hu2018federated} proposes a novel framework for environmental monitoring based on Federated Region Learning (FRL), designed to address the challenges associated with the exchange of monitoring data. Similarly, study \cite{han2019visual} applies FL in industrial manufacture visual inspection task to detect defects in production tasks and guarantee the privacy for manufacturers. Third, FL has also been introduced in healthcare domain \cite{kim2017federated, huang2019patient} to to break down the barriers of analysis throughout different hospitals and provide disease prediction for patients. For example, the study \cite{kim2017federated} employs tensor factorization models within a federated learning framework for phenotyping analysis, extracting valuable insights from health records while preserving patient confidentiality by avoiding the sharing of patient-level data.

\subsection{Poisoning Attacks on Federated Learning}

Due to the distributed nature, Federated Learning is vulnerable to various poisoning attacks by malicious clients and untrusted participants \cite{kairouz2021advances, mothukuri2021survey}. These poisoning attacks include model poisoning attacks and data poisoning attacks which seek to damage the global model by directly crafting the gradients or local training data respectively. 

Model poisoning attacks have been widely studied by researchers \cite{cao2022mpaf, fang2020local, baruch2019little,xie2020fall, damaskinos2019aggregathor, el2022genuinely, li2022learning}. To effectively perform such attacks, on the one hand, the disturbance should be strong enough to impact the behavior of the global model; on the other hand,  the crafted gradients should keep close to the benign gradients and avoid being noticed by the server and excluded (or reduced weight) from the aggregation. Reverse attack \cite{damaskinos2019aggregathor} first generates the benign gradient base on the training data owned, then the adversary crafts benign gradients by reversing the benign gradients. Random attack \cite{el2022genuinely} and Partial drop attack \cite{el2022genuinely} utilize a preset parameter $q\%$ to replace the $q\%$ original benign gradients parameters as a random value or 0 (drop the value) to generate the adversarial gradients. To keep close to the benign gradients, Little is enough attack \cite{baruch2019little} and Fall of empires attack \cite{xie2020fall} leverage the difference that exist across benign gradients. Specifically, the adversary first generates the benign gradients on all controlled clients and then adds perturbation to the mean of the gradients. Similarly, Local model poisoning attack \cite{fang2020local} leverages the statistical value of gradients owned. Based on the generated gradients and the global historical model update, Local model poisoning attack infers the convergence direction of the gradients and uploads the scaled, reverse gradient to poison the global model. Different from relying on the fixed attack strategy, \cite{li2022learning} achieves a stronger adaptive attack by leveraging the power of model-based reinforcement learning. In particular, the malicious participants first estimate the aggregated data distribution and consequently build a FL environment simulator based on the estimation to perform and support the adaptive attack.

Instead of directly adding perturbation on the gradient, the data poisoning attack \cite{zhang2022neurotoxin, lewis2023attacks, tolpegin2020data, goodfellow2014explaining} generates the crafted gradient through poisoning training data. On the one hand, some adversaries \cite{zhang2022neurotoxin, goodfellow2014explaining} aim to use crafted gradients to implant so called “backdoor” into the global model. As a result, the learned model’s outputs will be fixed to a preset target when facing certain inputs. These implanted backdoors could be semantics or graphics, for instance \cite{zhang2022neurotoxin} forces the model to generate the response “{RACE} people are psycho” when the user type “{RACE} people are”,  \cite{goodfellow2014explaining} can implant an imperceptible backdoor on the GoogLeNet \cite{szegedy2015going}, which leads the model to classify a panda as a gibbon when encountering the trigger. On the other hand, the adversary may seek to degrade the global model performance only on the target class by flipping the training labels \cite{lewis2023attacks, tolpegin2020data}, which is the so-called “label flipping attack”. When performing such attack, the adversary decides the attack strategy which includes the attacked class, the destination class, the amount of attacked classes, and the attacking timing. Then, the adversary flips the training label following the attack strategy and uses the poisoning data to train the local model. Once the poisoned gradient continuously participates in the aggregation, the global model will misbehave on the attacked class. As “label flipping attack” does not directly manipulate the gradient, it introduces a minor perturbation which makes it challenging to identify and defend. 

Today, “label flipping attack” has been introduced in various real world applications and shows significant effectiveness. Study \cite{taheri2020defending} applies this attack within a malware detection task for Android platforms, introducing a Silhouette Clustering-based Label Flipping Attack (SCLFA). This approach involves attackers calculating the silhouette value of each data sample and selecting those with a negative value as candidates to generate polluted data through label flipping. The paper \cite{zhang2021label} demonstrates that the “label flipping attack” also shows the effectiveness for fooling spam filtering systems. Through performing entropy method based flipping attacks, study \cite{zhang2021label} increases the false negative rate of Naive Bayes under the influence of label noise without affecting normal mail classification. On the other hand, the paper \cite{sharma2022catboost} introduces "label flipping attacks" in hardware Trojan detection systems. It demonstrates that model performance can be effectively deteriorated through a proposed stochastic hill-climbing search-based flipping attack, with only a small cost associated with flipping the labels. Given the substantial security challenges posed by "label flipping attacks", we focus on the these attacks \cite{lewis2023attacks, tolpegin2020data} in this work.

\subsection{Byzantine Resilient Federated Learning}
To defend against various poisoning attacks and enhance the robustness of federated learning, Byzantine resilient FL methods have been proposed by recent researchers \cite{li2022enhancing,cao2020fltrust,yin2018byzantine,blanchard2017machine}. Some of these works consider excluding the suspicious gradients from the aggregation base on the statistics. In particular, Krum \cite{blanchard2017machine} calculates the sum Euclidean distance (L2 distance) for a preset number of received gradients. The client that achieves the shortest distance is regarded as the most benign client, its gradient will become the new global gradient in this iteration. Trimmed Mean \cite{yin2018byzantine} and Median \cite{yin2018byzantine} are coordinate-wise aggregation methods that aggregate each parameter of the model, respectively. For each parameter, the server firstly sorts all the parameters received from client model updates. Then, the server trims the largest and smallest $p\%$ parameters and averages/takes the median value of the remaining gradients as the corresponding parameter of the new global model. FL Trust \cite{cao2020fltrust} enhances the FL robustness from the gradients' directions and magnitudes perspectives. Under the FL Trust framework, the server should collect a small clean dataset and keeps a corresponding model. In each learning round, FL Trust calculates the cosine similarity between the owned and clients' gradients and assigns a higher honest score for those clients achieving high cosine similarity. The clients' gradients are normalized by the server gradient and consequently participate in the aggregation; the client with a higher honest score will receive a heavier weight and vice versa. 

We note that most of these Byzantine resilient aggregation and the corresponding federated learning methods have been introduced to mitigate the model poisoning attacks in IID settings; the effectiveness when defending against data poisoning, especially in Non-IID scenarios still remains unclear.

\section{Motivation}

\subsection{Background and Definition}
Our problem is formulated on the synchronous federated learning (binary or multi-class) classifier task with a deep neural network (DNN) model. The DNN model consists of multiple layers, which are comprised of a corresponding set of nodes. During the learning process, each layer processes the input and sends the output to the following layers, where the first layer receives the training data as input, and the final layer generates the prediction result. FL paradigm enables multi-participants\footnote{In this work, we combined use “participants” and “clients” with the same meaning.} collaboratively train a model on their local devices without sharing their raw data. Specifically, in each FL learning iteration, the server first broadcasts the current global model $M$ to $n$ clients. Then, each client uses the data owned to locally trains the model $M$, and sends the gradient $g$ back to the server. Finally, the server aggregates all the model updates received by an aggregation method $\mathcal{A}(\cdot)$. The learning task starts from the server initializing the model and ends up with the global model $M$ achieving a preset accuracy. 

We use a joint distribution $p_i(x,y)$ to denote the private data owned by the $i$th participant, where $x = (x_1, ..., x_s)$ and $y = (y_1, ..., y_s)$ denote the feature space and the label space respectively; the label $y_s \in C$, where $C$ is the set of all possible classes. At the end of each learning iteration, the optimal model $m^*_i$ of client $n$ is the solution for the following optimization problem:

\begin{equation}
m^*_i = argmin_{m_i} F(m_i) \label{1}
\end{equation}
Here, $F(\cdot)$ is the expectation function:
\begin{equation}
F(m_i)=\mathbb{E}_{(x,y) \sim p_i}[l((x, m_i), y)] \label{2}
\end{equation}
where $l((x, m_i),y)$ represents the empirical loss. The joint distribution $p_i(x,y)$ could be composed as the following equation.
\begin{equation}
p_i(x, y)=p_i(x|y)p_i(y) \label{3}
\end{equation}

Here, $p_i(x|y)$ denotes the conditional distribution of feature $x$ given label $y$, and $p_i(y)$ represents the marginal distribution of label $y$ on the participant $i$. We note that in the real-world FL scenario, both the conditional distribution $p(x|y)$ and the marginal distribution $p(y)$ can be different across the clients due to its Non-IID nature. On the one hand, different clients may have various representations under a same label (i.e., $p(x|y)$). On the other hand, clients may own different amounts of data for each class which draws the heterogeneity of the marginal distribution (i.e., $p(y)$).

The differences between the local and optimal model of each client are subsequently calculated as the gradient $g$ and sent back to the server. The server aggregates the gradients through the aggregation rule $\mathcal{A}(\cdot)$ once the preset amount of $g$ have been received and subsequently uses it to update the global model, 
\begin{equation}
\begin{split}
g_i & = m^*_i -m_i \\
M^{t+1} = M^t + &\eta \cdot \mathcal{A}(g^t_i), \ i=1,2,...,n \label{4}
\end{split}
\end{equation}
where $t$ denotes the iteration and $\eta$ represents the learning rate which is a hyper-parameter. 

During the learning process, clients may craft and upload a malicious $g$ to poison the global model once penetrated by attackers. The attacker can directly manipulate the gradient $g$ or poison the local training data by mismatching the feature $x$ and the ground truth $y$ pairs to generate the crafted gradient. To defend against such attacks, some Byzantine-robust aggregation rules ($\mathcal{A}(\cdot)$) have been introduced. Instead of averaging the gradients received, these aggregation rules assign heavy weights to those gradients who behave honestly and vice versa.

In this paper, we focus on the label-flipping attacks (i.e., generate $g$ by mismatching $x$ and $y$) with different mismatching strategies, and agree that the marginal distribution heterogeneity is the main resource of Non-IID in FL. 

\subsection{Effectiveness of the existing Byzantine-robust aggregation rules}
In this section, we evaluate the effectiveness of the existing Byzantine-robust aggregation rules when defending against the label flipping attacks through MNIST in both IID and Non-IID settings. 
\subsubsection{Experimental Setup}
\textbf{Threat Model:} We assume 20 clients collaborate in the learning task to train a joint model $M$ and $25\%$ participants are controlled by the adversary which aims to decrease the testing accuracy of target class(es) $y$ on the global model $M$ through uploading crafted gradients. We consider the adversary can access the dataset on the controlled device, belonging to the original client, and utilize the dataset to generate the crafted gradient through the “label flipping attack” strategy. Under the proposed data distribution and learning scenario, the class 5 of MNIST receives the lowest testing accuracy in non adversarial setting (around $60\%$, shown in next section), and has been widely recognized as class 8 by the model, we regard it as the most vulnerable class and set the attack strategy as “flipping the training data label 5 as 8 on the attackers' device” to generate the crafted gradients in the proposed example.

\textbf{Training Data:} The MNIST database (Modified National Institute of Standards and Technology database) \cite{deng2012mnist} is an extensive handwritten digits collection that includes a training set of 60,000 examples and a test set of 10,000 examples; each example has consisted of $28\times28$ pixels and a label of $0\sim9$ to indicate the class belonging to. Under IID setting, we consider each client owns all training classes and evenly distribute the training data to all clients; we use $MNIST_{IID}$ to denote the case that MNIST training data with proposed IID setting. In Non-IID setting, we consider the most general scenario that clients' data may have different marginal distributions ($p(y)$) because of the interest heterogeneity and leverage the parameter $p$ to denote the Non-IID degree. During data distribution, we assign the $i$ client $p * s$ volume training examples with label $y$\footnote{The $y$ could be a single element or a set; in other words, the data could bias on one label or several labels.}, where  $s$ is the preset training data volume of each participant. Then, we evenly assign other class training data to client $i$ to fill up the rest $s - (p * s)$ capacity. We note that the client's data will heavily bias on the label $y$ if $p>\frac{1}{|U|}$, and only include the class $y$ once $p=1.0$, where $U$ is the labels universe of the learning task, when $y$ is a single label. In the proposed example, we set $P=0.9$ and use $MNIST_{0.9}$ to represent such case. 

\begin{figure}
\centering\centering
{\includegraphics [width=1.0\textwidth] {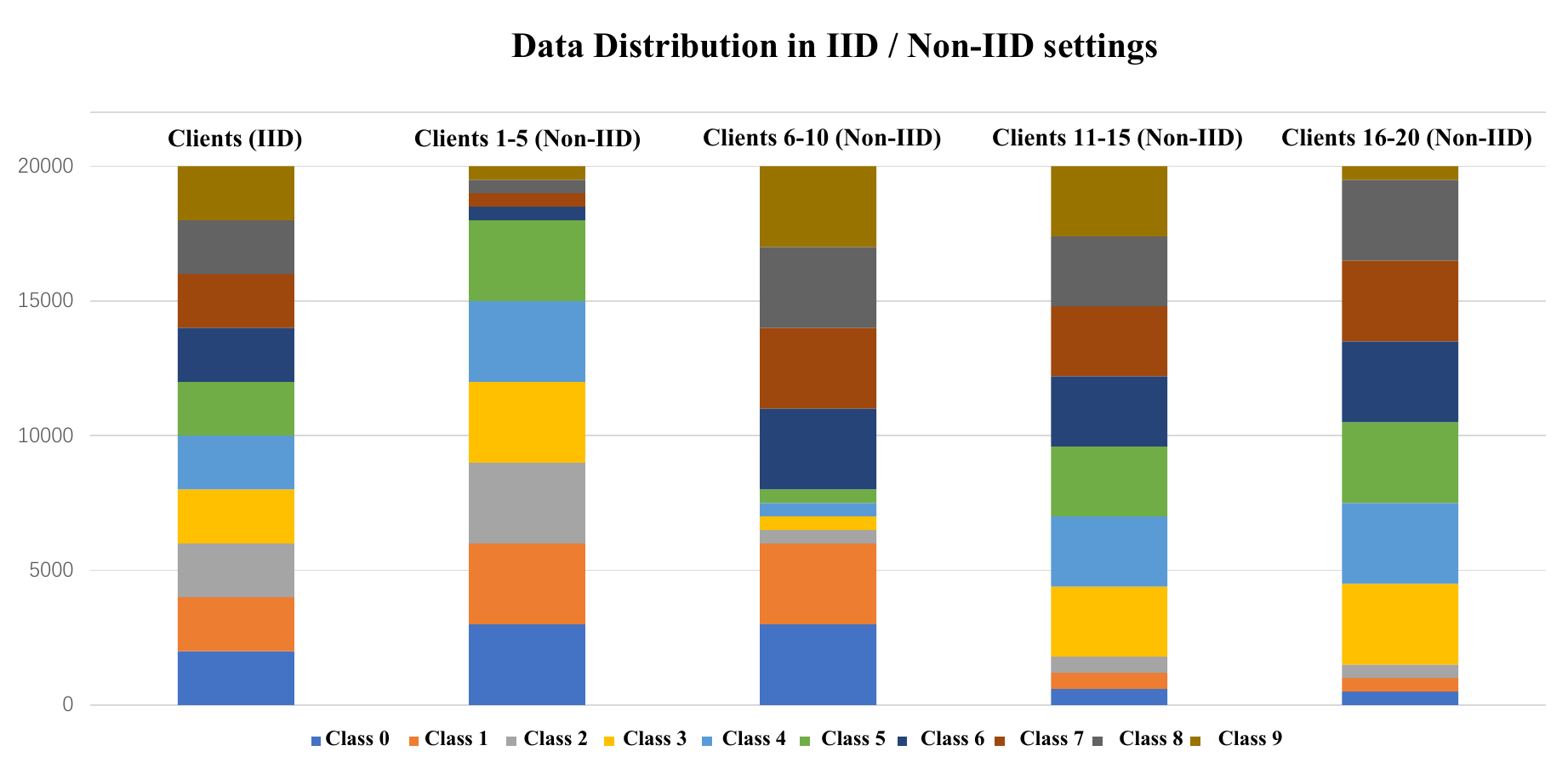}}
\caption{Illustration of the data distribution (IID:left bar, Non-IID right four bars). The bar indicates the training data owned by each client (client ID shown on the top of each bar), the colors indicate different classes of training data.}
\label{fig1}
\end{figure}

Figure \ref{fig1} illustrates the data distribution for 20 clients in IID and Non-IID (p=0.9) settings. Under the IID setting, all clients are evenly assigned each class data (left bar); under the Non-IID setting, the clients' data bias on different classes, making up four different distributions (right four bars). We assume $Client_{1}\sim Client_{5}$ play the attacker role.

\textbf{Training Model:} In this study, we use a general model for training, which includes a 28 $\times$ 28 dense layer and a  10$\times$ softmax layer. We set the learning rate as 0.001 and batch size as 128 to train the model 100 iterations.

\textbf{Benchmarks and Metrics:} As the “label flipping attack” \cite{lewis2023attacks, tolpegin2020data} aims to reduce accuracy of the targeted class, we use the accuracy as the metric, including global accuracy and the accuracy of each class. We select several representative byzantine robust aggregation FL methods (Krum \cite{blanchard2017machine}, Median \cite{yin2018byzantine}, Trimmed Mean \cite{yin2018byzantine} and FL Trust \cite{cao2020fltrust}) and evaluate their performance when facing label flipping attack. We further use the Vanilla FL \cite{mcmahan2017communication} in non-adversarial as the baseline in the IID setting to compare with. We have not include Vanilla FL in Non-IID setting as it is not been designed for such scenarios. In Non-IID setting, each FL method is compared with its performance achieved in IID settings.

\subsubsection{Experimental Results}
The results show the existing FL methods (1) can maintain the robustness when defense against the “label flipping attack” in IID scenarios (2) can not effectively aggregate the information of each client in Non-IID scenarios (3) can not defense against the “label flipping attack” in Non-IID scenarios.

\textbf{(1) The existing FL methods can maintain the robustness when defense against the “label flipping attack” in IID scenarios.} First, the existing FL methods achieve a similar global accuracy as Vanilla FL (Baseline). For instance, Vanilla FL achieves 84.5\% global accuracy while Krum, Median, Trimmed Mean and FL Trust achieve 84.43\%, 85.22\%, 84.47\% and 84.25\% respectively. On the other hand, they maintain the accuracy of the attacked class (i.e., class 5). Specially, class 5 receives 57\% testing accuracy under Vanilla FL and receives 56\%, 60\%, 57\% and 53\% in Krum, Median, Trimmed Mean and FL Trust, respectively. Besides, we note that the information of all classes have effectively learned by the global model through the existing FL methods which receives a similar accuracy compared with the Baseline.

\textbf{(2) The existing FL methods can not effectively aggregate the information of each client in Non-IID scenarios.} The result shows that the existing FL methods drop the global accuracy in Non-IID scenarios against “label flipping attack”. Specifically, Krum, Median, Trimmed Mean and FL Trust decrease the global accuracy from around 85\% to 65\%, 65\%, 67\% and 72\%, respectively. Although some classes have not been attacked, they still decrease testing accuracy (even to 0\%), which consequently cause the significantly global accuracy decrease. For instance, class 8 and 9 receive 0\% in Krum and Median FL method; receive 24\% and 0\% under Trimmed Mean. In contrast, FL Trust maintains a relatively stable accuracy across different classes, only class 8 witnesses a significant accuracy reduce.

\textbf{(3) The existing FL methods can not defend against the “label flipping attack” in Non-IID scenarios} The existing FL methods failed to maintain the accuracy of attacked class in Non-IID settings. Due to the “label flipping attack ”, class 5 testing accuracy drops to 24\% in Trimmed Mean and to 0\% in other three FL methods.   

\begin{figure}[h]
\centering
{\includegraphics [width=0.48\textwidth] {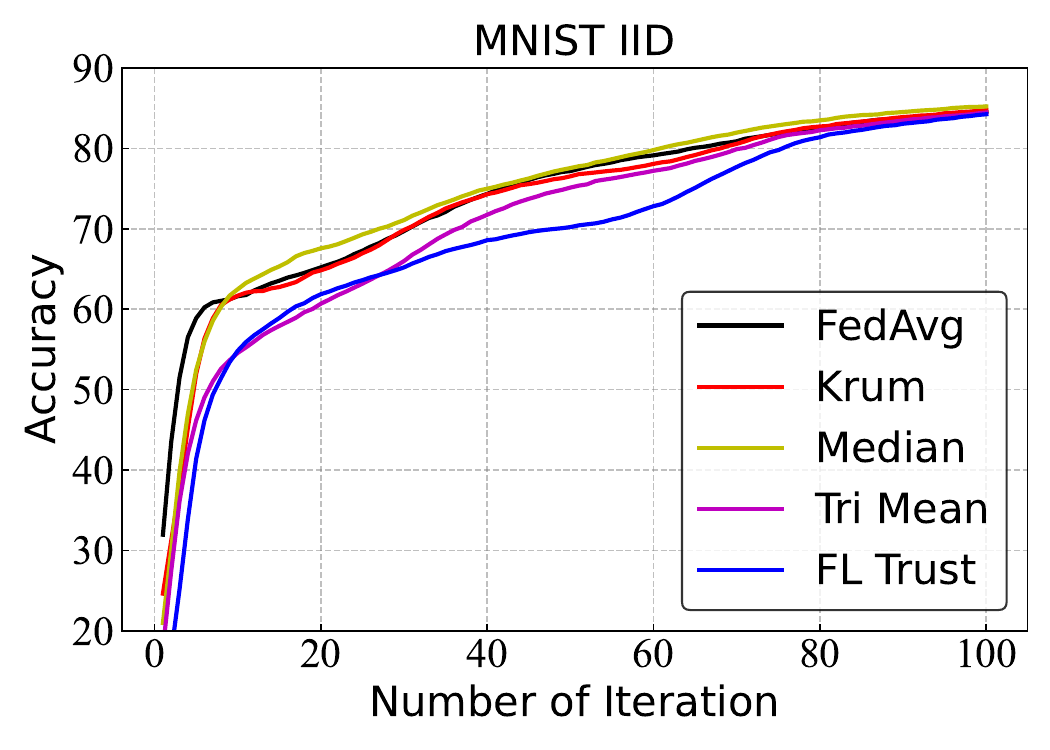}}
{\includegraphics [width=0.48\textwidth] {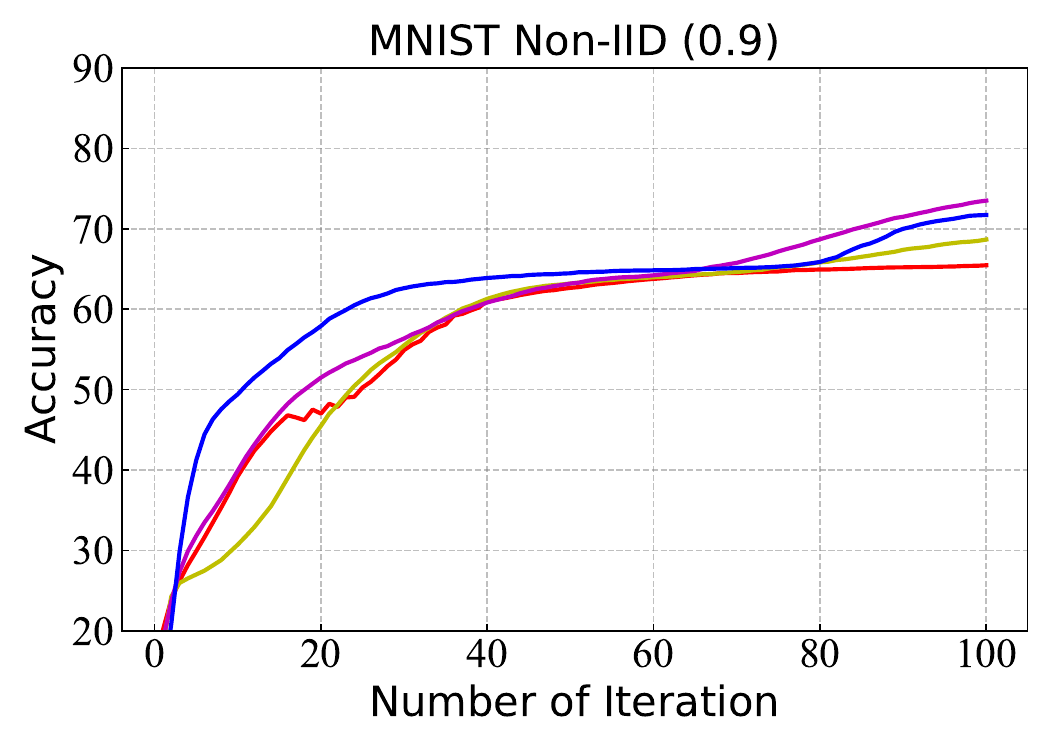}}
\caption{Illustration of the global accuracy of the existing FL methods against “label flipping attack” in IID and Non-IID scenarios}
\label{fig2}
\end{figure}

Figure \ref{fig2} and Table \ref{tab1} illustrate the global accuracy and accuracy of different classes of different FL methods against “label flipping attack” in IID and Non-IID scenarios, respectively. In next section, we will provide detailed discussion for the reason pertaining to observed phenomena (2) and (3).

\begin{table}
\caption{Illustration of the classes' accuracy of the existing FL methods against “label flipping attack” in IID and Non-IID scenarios}
\begin{center}
\begin{tabular}{c|c|c|c|c|c|c|c|c|c|c}
\toprule[1pt]
\midrule
\textbf{Class}&\textbf{0}&\textbf{1}&\textbf{2}&\textbf{3}&\textbf{4}&\textbf{5}&\textbf{6}&\textbf{7}&\textbf{8}&\textbf{9}\\
\midrule
\textbf{FedAvg}&96\%&95\%&89\%&88\%&85\%&60\%&92\%&87\%&79\%&82\%\\
\midrule
\multicolumn{11}{c}{\textbf{$MNIST_{IID}$}}\\
\midrule 
\textbf{Krum}&96\%&97\%&81\%&87\%&84\%&56\%&92\%&86\%&78\%&83\%\\
\midrule
\textbf{Median}&97\%&96\%&81\%&86\%&85\%&60\%&92\%&85\%&79\%&93\%\\
\midrule
\textbf{Tri Mean}&96\%&96\%&81\%&87\%&77\%&57\%&92\%&86\%&81\%&83\%\\
\midrule
\textbf{FL Trust}&95\%&96\%&75\%&88\%&81\%&53\%&91\%&86\%&81\%&93\%\\
\midrule
\multicolumn{11}{c}{\textbf{$MNIST_{0.9}$}}\\
\midrule 
\textbf{Krum}&95\%&97\%&83\%&92\%&94\%&0\%&92\%&81\%&0\%&0\%\\
\midrule
\textbf{Median}&92\%&96\%&80\%&92\%&94\%&2\%&95\%&87\%&0\%&0\%\\
\midrule
\textbf{Tri Mean}&93\%&96\%&78\%&92\%&93\%&0\%&94\%&88\%&24\%&0\%\\
\midrule
\textbf{FL Trust}&95\%&93\%&85\%&89\%&86\%&0\%&93\%&86\%&0\%&77\%\\
\midrule
\bottomrule[1pt]
\end{tabular}
\label{tab1}
\end{center}
\end{table}

\subsubsection{Discussion}
In this section, we provide the discussion and answers of two research questions based on the experimental result. (Q1) Why do the existing FL methods fail to aggregate the information of different clients under Non-IID scenarios? (Q2) Why do the existing FL methods fail to defend “label flipping attack”?

\textbf{A1:} First, Krum \cite{blanchard2017machine}, and Median methods \cite{yin2018byzantine} enhance model robustness by selecting the most “honest” gradient (the shortest L2 distance) or coordinate-wise (median value) parameter as the aggregation outcome and updating the global model. As the client owns class-balanced training data in the IID setting proposed, one client gradient can evenly carry all class information which enables the global model to increase accuracy and collect information on each class. However, in Non-IID settings, especially when the class imbalance exists, the client gradient may be biased and majorly carry information of the biased class. Directly using one client’s gradient as the aggregation outcome in such cases may lead the global model to achieve good learning performance on the biased classes but put the minor classes (e.g., Class 8, 9 in the example) at disadvantages. Second, the Trimmed Mean \cite{yin2018byzantine} enhances the model robustness by trimming the extremum of each parameter. Unlike model poisoning attacks that directly scale each parameter, the “label flipping attack” generates the crafted gradient through poisoning training data. As the attack strategy introduces a smaller crafting magnitude (compared with model poisoning attack), trimming extremum may not effectively remove adversarial parameters but remove the parameter from benign clients instead, which leads the global model to lose information (i.e., Class 2 and 9) in aggregation and achieve a decreased accuracy. Besides, these methods are relatively fixed when generating the aggregation and assigning weights. For instance, the cosine similarity between client's and severe gradient is highly related to the data distribution and quality of the client. The client owning IID-like and high-quality training data can continuously receive a high cosine similarity and heavyweight in FL Trust \cite{cao2020fltrust} aggregation unless it changes the data distribution pattern. In contrast, other clients will hardly participate in the aggregation (or receive a small weight), putting the global model in the difficulty that collecting information from these clients. 

\textbf{A2:} To achieve “label flipping attack” and effectively reduce the accuracy of the targeted class, the malicious clients (and the corresponding gradients) should keep participating in the learning task; once absent, the reduced accuracy will quickly return to the baseline \cite{tolpegin2020data}. In other words, in the proposed example, the attackers are excluded from the aggregation in the IID setting but can keep participating in the learning task of the Non-IID setting under the existing FL methods. We consider this because even the benign gradients are relatively heterogeneous in the Non-IID setting which drives difficulty in identifying the crafted gradient. Specifically, the existing FL methods regard the outliers (gradients or parameters) as adversarial and enhance model robustness by excluding these potential malicious gradients or parameters from the aggregation. In IID scenarios, clients' gradients tend to be similar, the scaled gradient can be easily identified and excluded by the existing FL methods. In contrast, even benign clients' gradients could be very different in the Non-IID setting. As the gradient generated by “label flipping attack” caries the right information of most (uncrafted) classes, it can mingle in the benign gradients which brings difficulty for the existing FL methods.

\begin{figure}
\centering
{\includegraphics [width=0.48\textwidth] {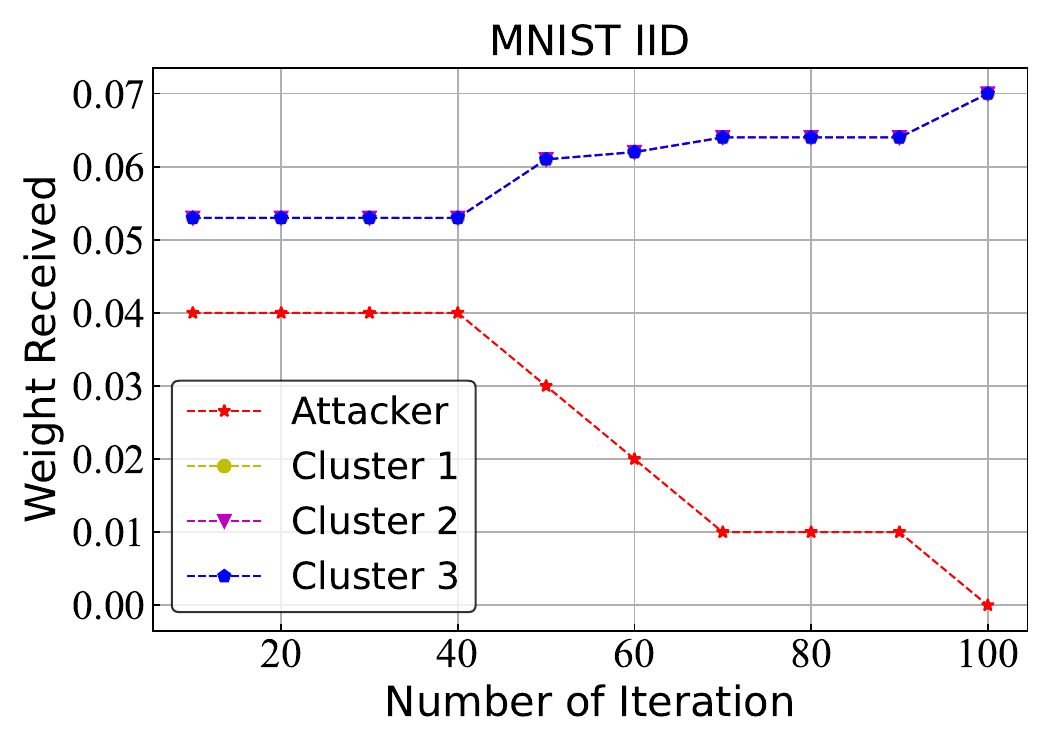}}
{\includegraphics [width=0.48\textwidth] {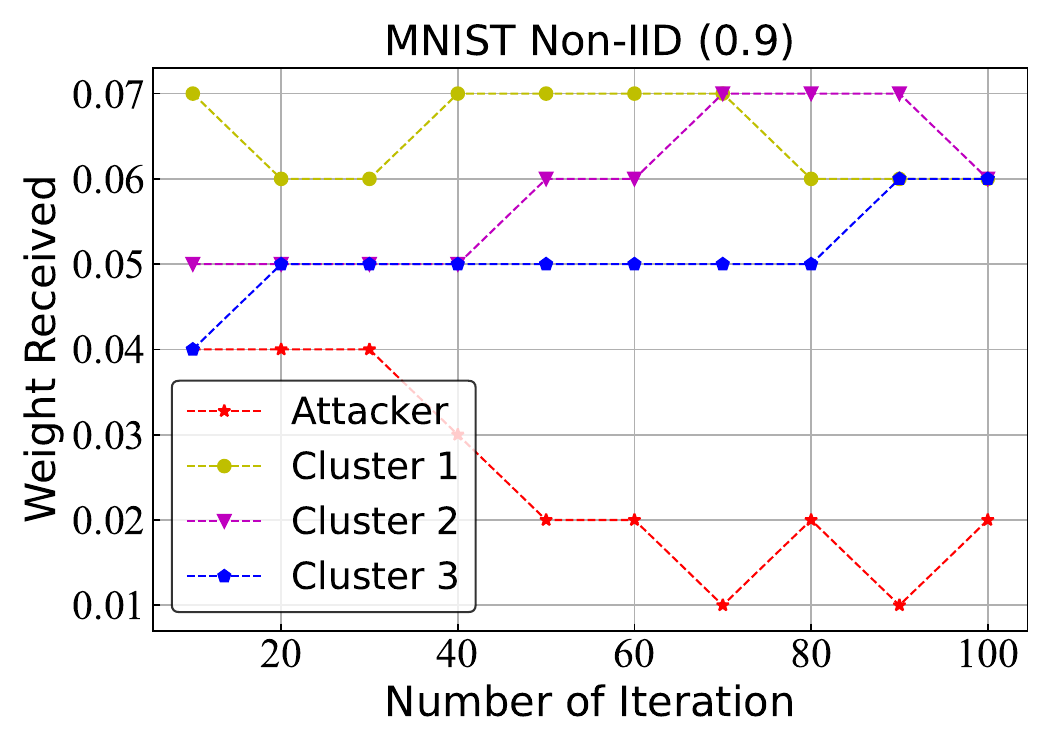}}
\caption{Illustration of the weight assigned by each client under the existing FL methods against “label flipping attack” in IID and Non-IID scenarios per 10 iterations}
\label{fig3}
\end{figure}

We provide an example for monitoring the weight of each client cluster (refer to the data distribution in Section 3.2.1) per 10 iterations in aggregation and note that the crafted gradients can keep participating in the aggregation even within the existing FL Trust. Figure \ref{fig3} illustrates the weight assigned each client cluster in the FL Trust. As there are four different data distributions (four client clusters) in the Non-IID setting, we monitor the weight assigned by each client with different data distributions each 10 learning round. The results show that the attacker (client cluster 1) receives a decreasing aggregation weight within the whole learning task in the IID setting, and be excluded from aggregation since iteration 90 (i.e., receives weight 0\%). In contrast, attackers keep participating in the learning task although witnessing a decreasing weight; at the 100th iteration, each attacker still receives 2\% weight which makes up 10\% in total. Besides, we note that although client clusters 2, 3, and 4 are all benign, they receive unequal weight in the aggregation, which explains why Class 1 $\sim$ 10 have different learning performances. 

\section{Honest Score Client Selection based Federated Learning}
\subsection{Overview}
To enhance the robustness of FL system, we propose the Honest Score Client Selection scheme (short as HSCS) and corresponding FL framework (short as HSCSFL). Instead of eliminating the impact of the adversary during the aggregation, HSCSFL considers only selecting top $p\%$ contribution score clients to participate in the learning task and excludes the adversary before the aggregation; here $p\%$ is a preset parameter base on the knowledge. In the next section, we will discuss why the gradient generated by “label flipping attack” achieves a lower contribution score compared with the benign gradients and how does contribution score client selection scheme work. The objectives of HSCSFL are the following:

\begin{itemize}
\item HSCSFL should effectively aggregate the information of different classes in Non-IID scenarios; in other words, classes should achieve a similar learning performance.  
\item HSCSFL should be robust when defending against “label flipping attack”, the attacked class should keep the accuracy similar to the non-adversary level.     
\end{itemize}

\subsection{Assumption}
 In HSCSFL, we consider the server is honest and acts as a defender. The server should be able to collect a clean evaluation dataset that evenly includes each class. We keep the assumption set in Krum \cite{blanchard2017machine} that the server knows the number (i.e., $p\%$) of malicious clients. We consider both IID and Non-IID scenarios and agree on the marginal distribution deference is the main Non-IID resource. We follow a general setting in which all clients (i.e., benign and adversary clients) keep relatively similar diversity and volume training data.  
 
\subsection{Honest Score Client Selection scheme}
\subsubsection{Honest Score Client Selection scheme framework}
We consider a dynamic client selection scheme to select benign participants based on the current global model and exclude potential malicious clients before aggregation. As the key for attackers reducing target class accuracy through “label flipping attack” is participating in the aggregation each iteration \cite{tolpegin2020data}, the proposed scheme should notice immediately once the model has been attacked and exclude the potential adversary in the next few learning rounds. We consider the server plays the defender’s role and can collect a small, clean evaluation dataset to monitor the global model. As the “label flipping attack” only decreases the accuracy of the attacked class, evaluating the overall performance of client gradients is not enough and can not identify the crafted gradients effectively. Thus, our contribution score client selection evaluates each class separate ly and generates the corresponding accuracy vector. We further monitor the accuracy of each class of the global model through the evaluation dataset and generate the potential risk victor, the class that achieves a low accuracy will be assigned a high risk value. Finally, the server generates the honest score for all clients by calculating dot product of the clients' accuracy vector and the global model's risk vector, only top $p\%$ clients can participate in the following aggregation. By leveraging the risk vector, the server can effectively notice if a class has been attacked and flag the class with a high risk value. Adversaries that attend to attack that class will receive a lower honest score than other benign clients in the next learning round, which can not further participate in the aggregation. 

\begin{figure}
\centering
{\includegraphics [width=1.0\textwidth] {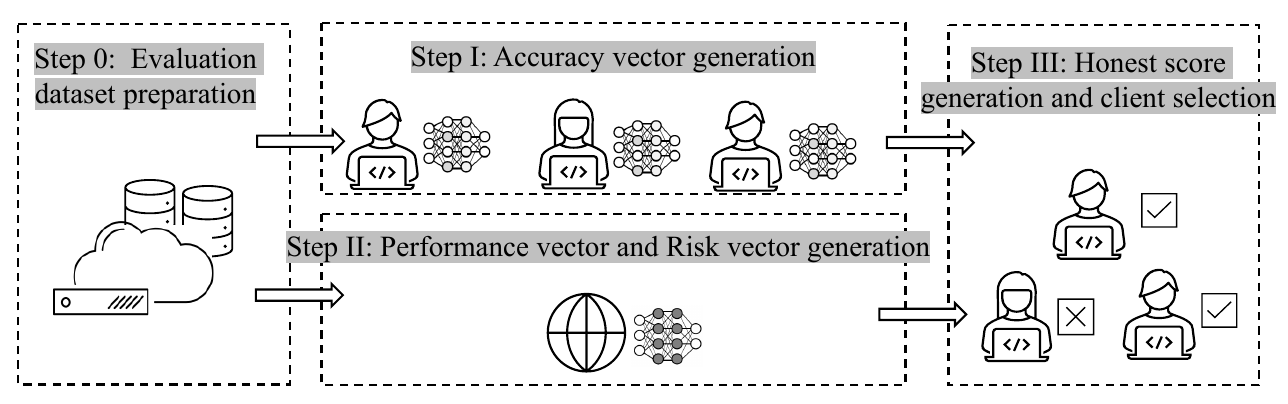}}
\caption{Illustration of the framework of the Honest Score Client Selection scheme}
\label{fig4}
\end{figure}

Specifically, honest score client selection includes the following four steps. Figure \ref{fig4} illustrates the framework of the Honest Score Client Selection scheme.
\begin{itemize}
\item Step 0: Evaluation dataset preparation

To identify the malicious gradients, HSCS requires the server to collect a small clean evaluation dataset before the learning task start. This evaluation dataset should evenly include each class data. In this study, we set the evaluation data size as 5\% of the overall training data volume. Follow the research \cite{cao2020fltrust, park2021sageflow}, we consider the evaluation dataset could be collected manually.
\item Step I: Accuracy vector generation

Once a preset number of clients' gradients have been received, the server evaluates the local model updates through the evaluation dataset. Different from FL Trust and Sageflow evaluate the local model updates from the overall perspective, HSCS evaluates the clients' gradients from each class. At the end of Step I, all local updates should be evaluated and each of them has an “accuracy vector” (represent as $AccV$) which records its performance (i.e., accuracy) on each class.  

\item Step II: Performance vector and Risk vector generation

To identify the class that potentially be attacked and avoid being further affected, we monitor the accuracy of each class of the global model and flag the lower accuracy class with a higher risk value. Specifically, the server first evaluates the global model at this round and generates the “global model performance vector” (represent as $PerV$)which includes each class's accuracy. Then, the server generates the “Risk vector” (represent as $RisV$) by the following equation:
\begin{equation}
RisV = 1- PerV \label{5}
\end{equation}
At the end of Step II, the class with high accuracy of the global model is assigned with a low risk value and vice versa; all risk values are recorded in the $RisV$.

\item Step III: Honest score generation and client selection

Finally, the server generates the honest score (represent as $HS$) for each client based on the $AccV$ and $RisV$ by equation \ref{6}, where $i$ indicates the $i$th client, $C$ denotes the number of class. $AccV$ and $RisV$ include $c$ elements respectively.

\begin{equation}
HS_i = \sum_{j=1}^{C}AccV_i * RisV \label{6}
\end{equation}

The server subsequently selects the top $p$ \% $HS$ clients to participate in the following aggregation. The $p\%$ is a preset parameter designed based on the knowledge of the amount of the adversary. 
\end{itemize}
\subsubsection{Toy example}
In this section, we provide a toy example to further demonstrate and clarify how the HSCS scheme work. We consider a simple 3 classes ($c = 3$) FL scenario with a server and 4 clients ($n = 4$). Suppose an evaluation dataset that evenly includes 3 classes has already been prepared. At the $t$th learning round, the server first evaluates 4 clients' gradients through the evaluation dataset and generates the corresponding $AccV$. Suppose the clients' $AccV$ is generated as the following:     

\begin{equation}
AccV_1 = [0.71, 0.82, 0.65], \  ...,\  AccV_4 = [0.41, 0.80, 0.97]
\end{equation}

Then, the server evaluates the current global model by the evaluation dataset to generate the $PerV$ and corresponding $RisV$. Suppose the $PerV$ and $RisV$ are generated as the following:     
\begin{equation}
PerV = [0.24, 0.55, 0.57], \  RisV = 1- Per = [0.76, 0.45, 0.43]
\end{equation}

The server generates the $HS$ for each client base on their $AccV$ and the $RisV$. Here, we demonstrate the process for generating $HS_1$:

\begin{equation}
HS_1 = \sum_{j=1}^{3}AccV_1 * RisV = 0.71*0.76 +0.82*0.45+0.65*0.43=1.19
\end{equation}

Finally, top $p\%$ HS clients will be selected and participate in the following aggregation. Suppose $p=50\%$, $Hs_1$ and $HS_2$ are the top two scores, client 1 and 2 will be consequently selected in aggregation.

This toy example also shows the importance and benefits of evaluating the performance of the client's gradient in each class. Specifically, class 1 achieves a significantly lower accuracy on the current global model; compared with other classes, class 1 is highly likely to be attacked. When evaluating clients from the overall perspective, client 1 and client 4 are difficult to be compared as they achieve a similar performance (mean accuracy as 72\%). However, as class 1 has a high risk of being attacked, Client 4 is more suspicious in the current iteration, which should be avoided to be selected. Thus, the HSCS scheme assigns client 4 a lower $HS$ than client 1, which can effectively exclude client 4 from the subsequent aggregation. In this toy example, we assume that the evaluation dataset is clean and unbiased. For scenarios involving a polluted dataset, we note that the correct $AccV$ can not be guaranteed and we provide a formal discussion in Section 6.2.

\subsection{Honest Score Client Selection scheme based Federated Learning}
\begin{figure}[h]
\centering
{\includegraphics [width=1.0\textwidth] {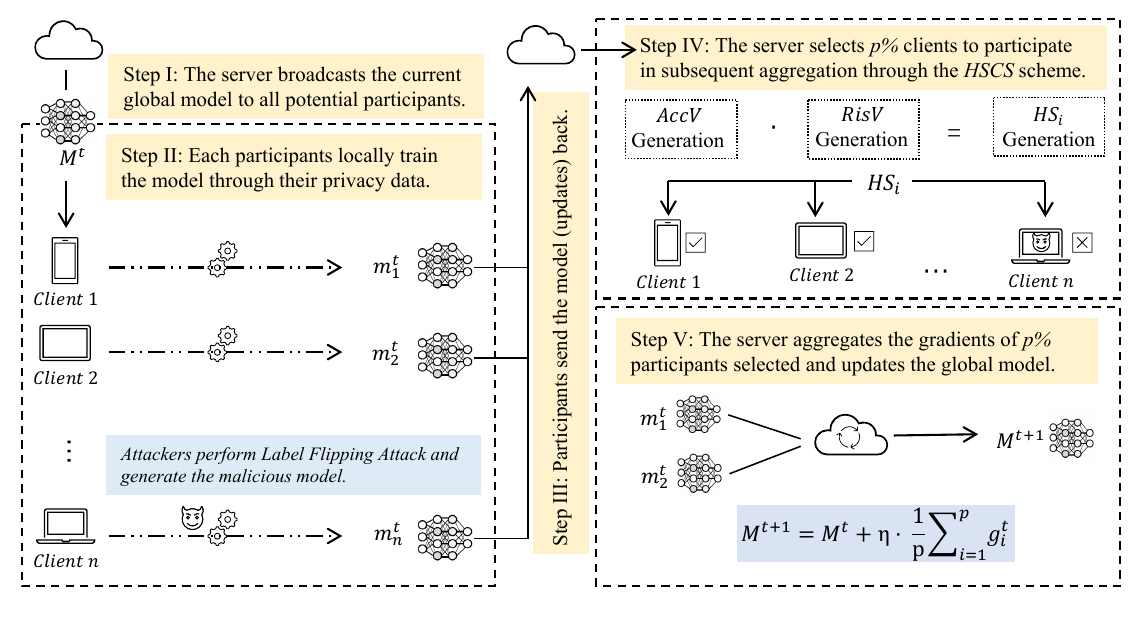}}
\caption{Illustration of the HSCSFL framework}
\label{fig5}
\end{figure}

The Honest Score Client Selection scheme based Federated Learning (short as HSCSFL) follows the standard (Vanilla) FL \cite{mcmahan2017communication} and introduces the HSCS before the aggregation stage to mitigate “label flipping attack.” Specifically, HSCSFL includes the following five stages, Figure \ref{fig5} illustrates the HSCSFL framework.
\begin{itemize}
\item Stage 1: At each learning iteration, the server first broadcasts the current global model to all participants and potential clients.
\item Stage 2: After receiving the current global model from the server, each client locally trains the model through the privacy data.
\item Stage 3: Once the preset model accuracy or training round reached, all clients send the local model updates back to the server.
\item Stage 4: Based on the clients' gradient received, the server performs HSCS scheme to select $p\%$ client to participate in subsequent aggregation. Specifically, the server first generates the Accuracy Vector for all clients and the Risk Vector of the current global model. Then, the server calculates and ranks the Honest Score for all participants to select aggregation candidates.
\item Stage 5: Finally, the server aggregates the $p\%$ gradients by averaging. The aggregation outcome is subsequently used to updates the global model to finish this learning iteration. Formula \ref{10} illustrates the aggregation rule of HSCSFL.
\end{itemize}
\begin{equation}
M^{t+1} = M^t + \eta \cdot \frac{1}{p}\sum^p_{i=1}{g^t_i}, \ p=p\% \cdot n \label{10}
\end{equation}

\section{Evaluation}
In this section, we comprehensively evaluate the robustness of HSCSFL when defending against “label flipping attack.” We consider different Non-IID degrees and attack strategies in the evaluation.

\subsection{Dataset and Model Sittings}\label{data}
Expect the MNIST dataset introduced in Section 3.2.1, we further introduce the Fashion MNIST dataset \cite{xiao2017fashion} for evaluation. Fashion MNIST (short as FMNIST) consists of a training set of 60,000 examples and a test set of 10,000 examples. Each example is a 28x28 gray-scale image associated with a label from 10 classes of fashion items. To simulate different Non-IID scenarios in the real world, we set the Non-IID degree as 0.8 and 0.9 to generate $MNIST_{0.8}$, $MNIST_{0.9}$, and $FMNIST_{0.8}$, $FMNIST_{0.9}$. We follow the data distribution introduced in Section 3.2.1 to distribute data to 20 clients. We consider the server collects/owns a clean evaluation dataset to support performing the HSCS scheme.

We follow the Section 3.2.1 and use a general model for training. This model includes a 28 $\times$ 28 dense layer and a  10$\times$ softmax layer. We train the model 100 iterations through setting the learning rate as 0.001 and batch size as 128.

\subsection{Benchmark}
We keep using the representative Byzantine resilient FL methods introduced in Section 3 as benchmarks. We also include the performance of FedAvg under IID setting in the experiments. However, the testing value of FedAvg only indicates the class or global accuracy's potential upper bound and can not directly compared with other benchmarks because of different training data distributions. We include FedAvg with IID scenario here to demonstrate our HSCSFL can reach the upper bound of learning performance in several cases.

\begin{table}
\caption{Illustration of the different attack strategies of “Label Flipping Attack.”}
\begin{center}
\begin{tabular}{c|c|c|c|c}
\toprule[1pt]
\midrule
\textbf{Strategy}&\textbf{Dataset}&\textbf{Adversary}&\textbf{Original Class}&\textbf{Flipping Class}\\
\midrule
1&MNIST&25\%&5&8\\
\midrule
2&Fashion M&25\%&5&8\\
\midrule
3&Fashion M&25\%&0, 1&8\\
\midrule
\bottomrule[1pt]
\end{tabular}
\label{tab2}
\end{center}
\end{table}

\subsection{Threat Model}
We consider $25\%$ clients have been controlled by the adversary; the attackers can generate the crafted gradient through the Label Flipping Attack based on the data owned. We consider the adversary can select different attacking strategies. Table \ref{tab2} illustrates the different attack strategies of the Label Flipping Attack simulated in the experiments. Specifically, we simulate the adversary attacks on the most vulnerable class (class 5 in MNIST), normal class (class 5 in FMNIST), and multiple classes (class 0 and 1 in FMNIST). 

We also consider an adaptive adversary environment, which includes 15\% and 35\% overall attackers. We leave the adaptive experiments and the corresponding discussion in subsection [LATER].

\begin{table}
\caption{Illustration of the global accuracy under different attack strategies of “Label Flipping Attack.”}
\begin{center}
\begin{tabular}{c|c|c|c|c|c}
\toprule[1pt]
\midrule
\textbf{Dataset}&\textbf{Krum }&\textbf{Median }&\textbf{Tri Mean }&\textbf{FL Trust}&\textbf{HSCSFL}\\
\midrule
\multicolumn{6}{c}{Atatck Strategy 1}\\
\midrule
$MNIST_{0.8}$&65.49\%&68.66\%&73.50\%&73.7.\%&87.16\%\\
\midrule
$MNIST_{0.9}$&65.04\%&65.71\%&67.22\%&72.45\%&88.25\%\\
\midrule
\multicolumn{6}{c}{Attack Strategy 2}\\
\midrule
$Fashion M_{0.8}$&67.59\%&62.13\%&70.51\%&69.11\%&76.73\%\\
\midrule
$Fashion M_{0.9}$&65.07\%&56.40\%&68.55\%&60.21\%&75.64\%\\
\midrule
\multicolumn{6}{c}{Attack Strategy 3}\\
\midrule
$Fashion M_{0.8}$&73.43\%&65.39\%&72.54\%&70.01\%&76.74\%\\
\midrule
$Fashion M_{0.9}$&72.67\%&62.59\%&66.61\%&59.31\%&77.20\%\\
\midrule
\bottomrule[1pt]
\end{tabular}
\label{Cab3}
\end{center}
\end{table}

\subsection{Experimental Results}
The experimental results show our HSCSFL effectively enhances the model robustness and maintains global accuracy when defending the Label Flipping Attacks with different attack strategies. Recall the objectives set in Section [SUB] are (I) effectively aggregate the information of different classes, and (II) effectively defend against Label Flipping Attacks in non-IID settings; our HSCSFL achieves both two objectives. Table \ref{tab3}, Figure \ref{CBfig6}, and \ref{fig7} compare the global accuracy of the different FL methods against Label Flipping Attacks in the FMINST and MNIST datasets with different non-IID scenarios.

\begin{figure}
\centering
{\includegraphics [width=0.48\textwidth] {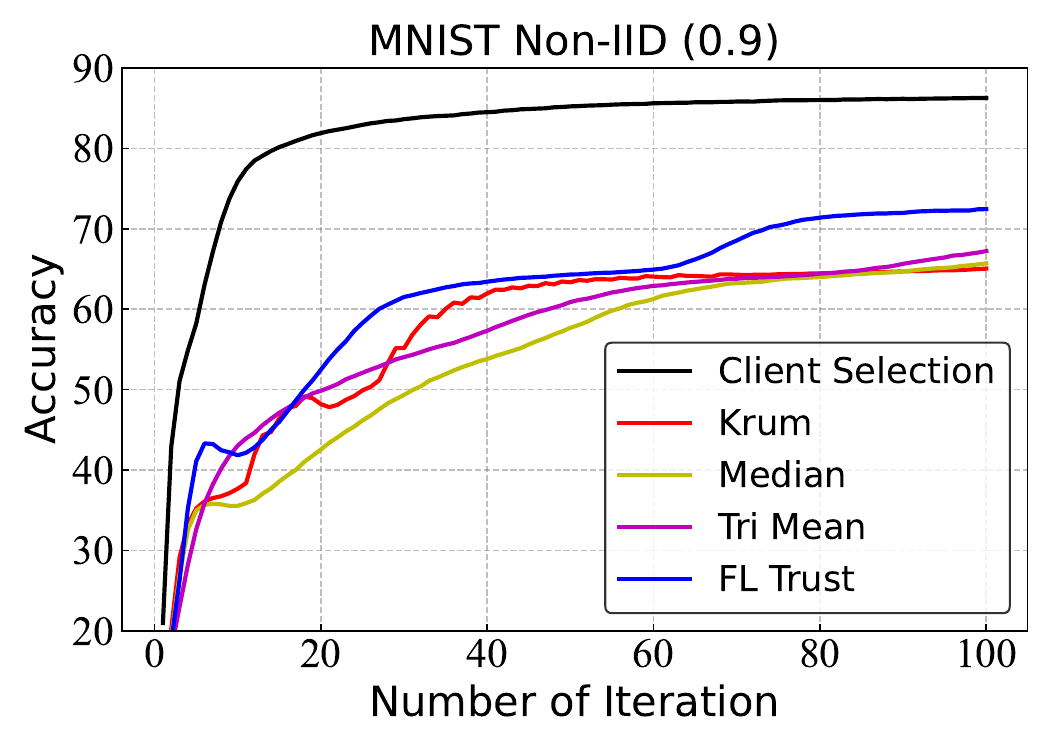}}
{\includegraphics [width=0.48\textwidth] {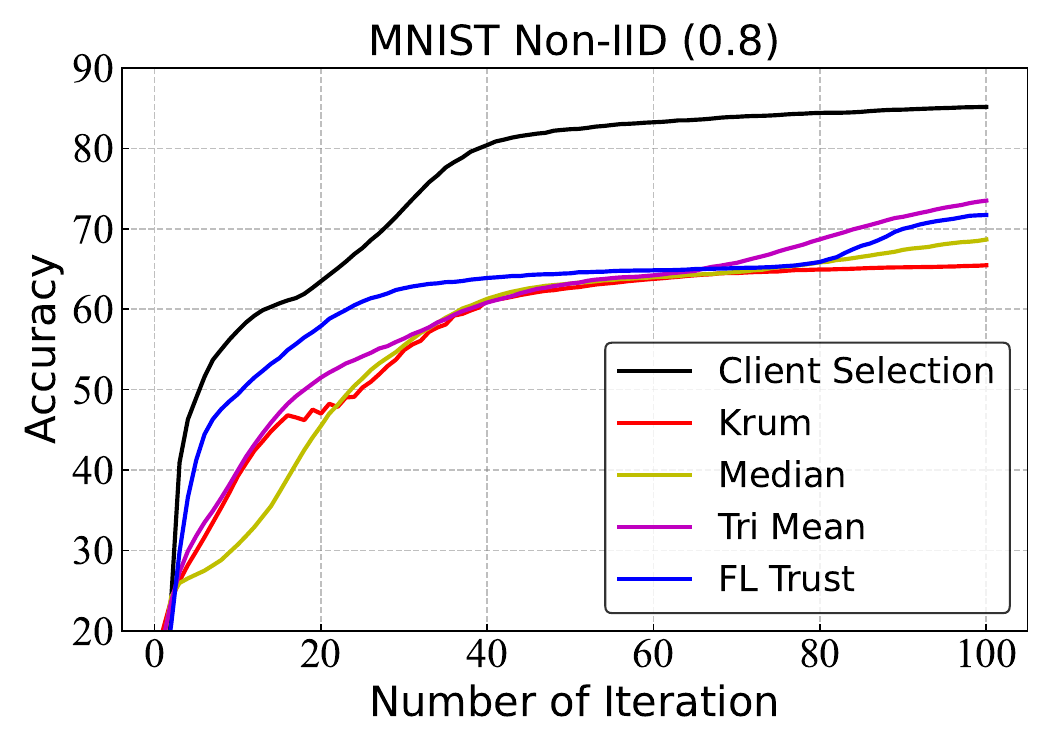}}
\caption{Illustration of the global accuracy of the different FL methods against “label flipping attack” (strategy 1) in MNIST with Non-IID scenarios}
\label{fig6}
\end{figure}

\begin{figure}
\centering
{\includegraphics [width=0.48\textwidth] {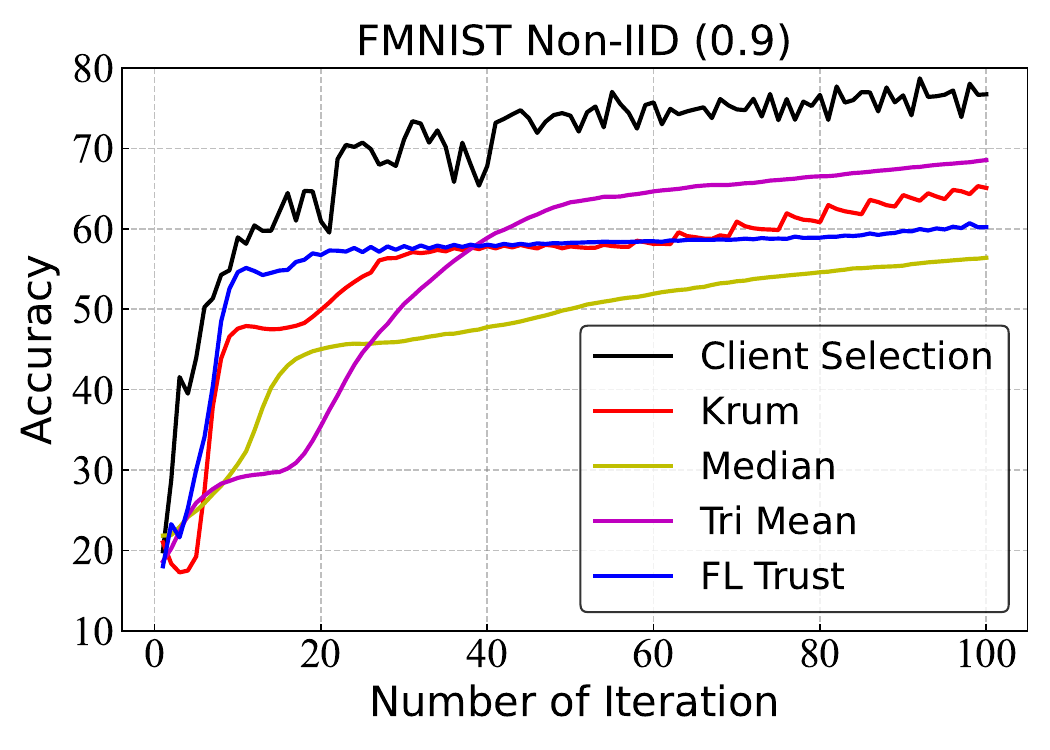}}
{\includegraphics [width=0.48\textwidth] {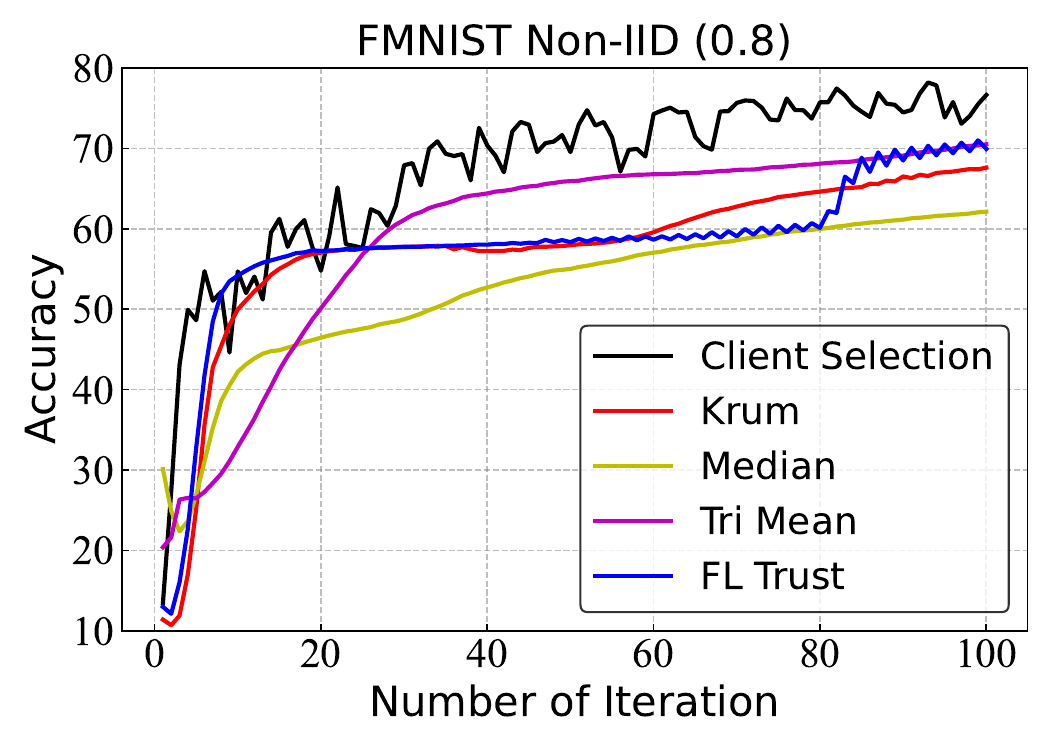}}
{\includegraphics [width=0.48\textwidth] {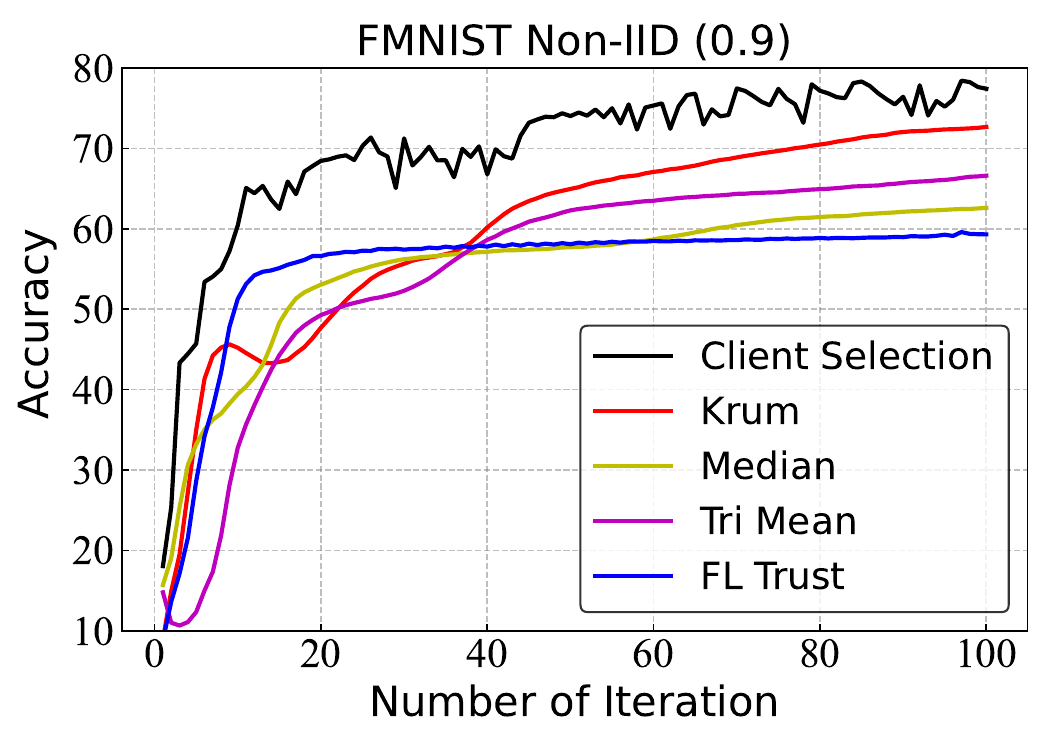}}
{\includegraphics [width=0.48\textwidth] {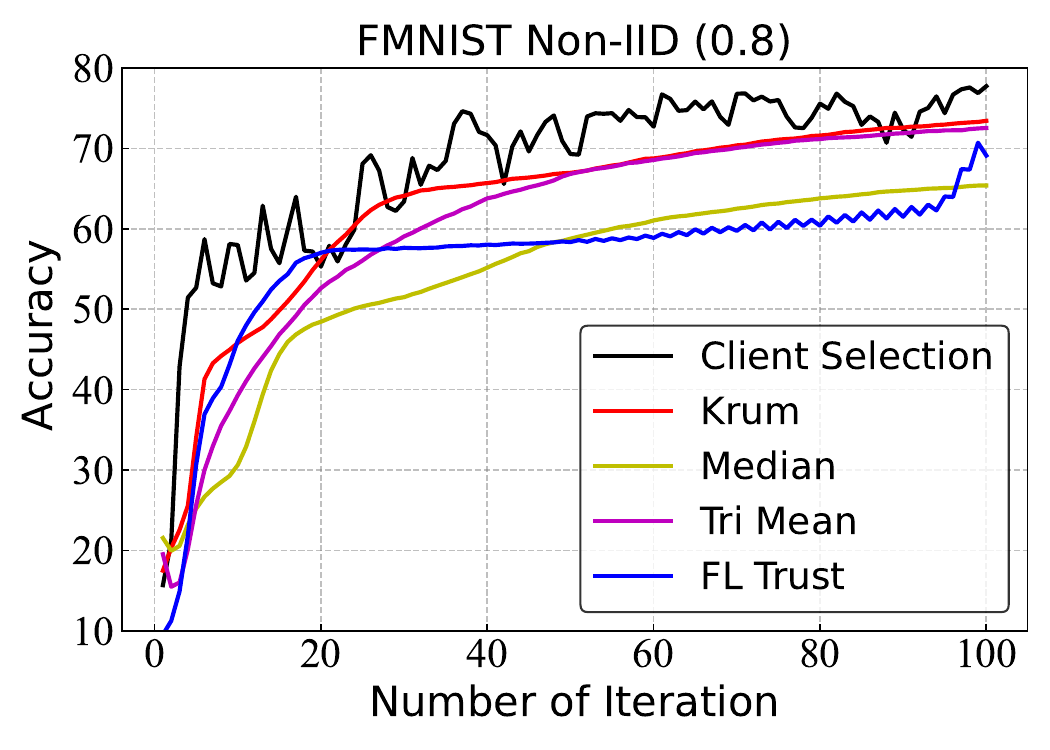}}
\caption{Illustration of the global accuracy of the different FL methods against “label flipping attack” with attack strategy 2 (up two), with strategy 3 (down two) in Fashion MNIST with Non-IID scenarios.}
\label{fig7}
\end{figure}
First, HSCSFL effectively aggregates the information of different classes in different non-IID settings and maintains global accuracy against the Label Flipping Attack. Vanilla FL \cite{mcmahan2017communication} achieves 84.5\% and 77.64\% in MNIST and FMNIST, while HSCSFL witnessed a similar testing accuracy, which achieves 87.16\%, 88.25\% in MNIST under attack strategy 1, 76.73\%, 75.64\% in FMNIST under attack strategy 2. When two FMNIST labels are attacked, HSCSFL still maintains a global accuracy of around 76.50\%. In contrast, the existing FL methods show a significantly degraded performance. For instance, Krum \cite{blanchard2017machine} and Median \cite{yin2018byzantine} decrease the global accuracy to around 65\% in two datasets when the non-IID degree is 0.8 and maintain the global accuracy when increasing the non-IID degree as 0.9. On the other hand, Trimmed Mean \cite{yin2018byzantine} and FL Trust \cite{cao2020fltrust} achieve a higher accuracy (70\%+) under two datasets when the non-IID degree is 0.8 but witness a decreasing model performance when increasing the non-IID degree. The global accuracy comparison shows our HSCSFL can effectively aggregate the client's model updates and learn information from different classes in various non-IID scenarios; we discuss the detailed accuracy of each class in the next paragraph. 

\begin{table}[h]
\caption{Illustration of the classes' accuracy of the existing FL methods against “Label Flipping Attack” in IID and non-IID scenarios.}
\begin{center}
\begin{tabular}{c|c|c|c|c|c|c|c|c|c|c}
\toprule[1pt]
\midrule
\multicolumn{11}{c}{\textbf{$MNIST_{0.9}$}}\\
\midrule
\textbf{Class}&\textbf{0}&\textbf{1}&\textbf{2}&\textbf{3}&\textbf{4}&\textbf{5}&\textbf{6}&\textbf{7}&\textbf{8}&\textbf{9}\\
\midrule
FedAvg&96\%&95\%&89\%&88\%&85\%&57\%&92\%&87\%&79\%&82\%\\
\midrule
HSCSFL&96\%&83\%&88\%&89\%&88\%&66\%&93\%&90\%&87\%&88\%\\
\midrule
\bottomrule[1pt]
\end{tabular}
\label{tab4}
\end{center}
\end{table}

Second, HSCSFL effectively defends against Label Flipping Attacks and maintains the attacked class accuracy in non-IID settings. Specifically, HSCSFL achieves 66\%, 65\%, 77\%, and 78\% accuracy on the attacked class in $MNIST_{0.9}$ and $MNIST_{0.8}$ under attack strategy 1, $FMNIST_{0.9}$ and $FMNIST_{0.8}$ under strategy 2 respectively; which is similar to, even higher to the upper bound carried by the FedAvg. On the contrary, almost all existing FL methods fail to maintain the model robustness and receive significant accuracy reduction on the attacked class. Krum shows the worst performance when defending against Label Flipping Attacks and receives 0\% class accuracy on almost all datasets. Under attack strategy 3, HSCSFL also achieves a good defending performance, which witnesses around 80\% and 92\% on two attacked classes.
\begin{table}
\caption{Illustration of the classes' accuracy of different FL methods against Label Flipping Attacks in MNIST with non-IID setting.}
\begin{center}
\begin{tabular}{c|c|c|c|c|c|c|c|c|c|c}
\toprule[1pt]
\midrule
\textbf{Class}&\textbf{0}&\textbf{1}&\textbf{2}&\textbf{3}&\textbf{4}&\textbf{5}&\textbf{6}&\textbf{7}&\textbf{8}&\textbf{9}\\
\midrule
\multicolumn{11}{c}{\textbf{$MNIST_{0.8}$}}\\
\midrule
FedAvg&96\%&95\%&89\%&88\%&85\%&57\%&92\%&87\%&79\%&82\%\\
\midrule
Krum&96\%&97\%&84\%&94\%&95\%&0\%&92\%&83\%&0\%&0.1\%\\
\midrule
Median&94\%&95\%&81\%&92\%&94\%&38\%&95\%&87\%&0\%&0\%\\
\midrule
Tri Mean&93\%&95\%&81\%&91\%&94\%&34\%&94\%&88\%&54\%&0\%\\
\midrule
FL Trust&95\%&95\%&84\%&90\%&89\%&0\%&94\%&88\%&0\%&65\%\\
\midrule
HSCSFL&97\%&97\%&85\%&88\%&92\%&65\%&93\%&89\%&79\%&80\%\\
\midrule
\bottomrule[1pt]
\end{tabular}
\label{tab5}
\end{center}
\end{table}
Besides, the unattacked classes have generally not been affected under HSCSFL. Unlike the existing FL methods tend to select one most honest client's gradient/parameter as the aggregation outcome, HSCSFL averages each client's gradient after performing the HSCS scheme, which enables their carried information can be equally learned by the global model. For instance, although class 8 has not been attacked in $FMNIST_{0.9}$, Krum \cite{blanchard2017machine}, Median \cite{yin2018byzantine}, Trimmed Mean \cite{yin2018byzantine}, and FL Trust \cite{cao2020fltrust} reduce the class accuracy to 55\%, 33\%, 84\%and 0\%, respectively; while HSCSFL achieves 90\% accuracy. Table \ref{tab4}, \ref{tab5}, and \ref{tab6} illustrate the detailed class accuracy of different FL methods under MNIST and FMNIST. For the sake of brevity, Table \ref{tab4} only includes the accuracy information of HSCSFL; the full experimental information can be found in Section [Motivation].

\begin{table}
\caption{Illustration of the classes' accuracy of different FL methods against Label Flipping Attacks in FMNIST non-IID settings.}
\begin{center}
\begin{tabular}{c|c|c|c|c|c|c|c|c|c|c}
\toprule[1pt]
\midrule
\textbf{Class}&\textbf{0}&\textbf{1}&\textbf{2}&\textbf{3}&\textbf{4}&\textbf{5}&\textbf{6}&\textbf{7}&\textbf{8}&\textbf{9}\\
\midrule
FadAvg&77\%&92\%&64\%&84\%&76\%&66\%&40\%&87\%&92\%&94\%\\
\midrule
\multicolumn{11}{c}{\textbf{$FMNIST_{0.9}$}}\\
\midrule
Krum&76\%&89\%&50\%&87\%&84\%&0.4\%&24\%&91\%&55\%&92\%\\
\midrule
Median&70\%&90\%&42\%&85\%&77\%&23\%&48\%&92\%&33\%&0\%\\
\midrule
Tri Mean&72\%&89\%&42\%&86\%&76\%&10\%&42\%&93\%&84\%&86\%\\
\midrule
FL Trust&83\%&91\%&68\%&85\%&78\%&11\%&0.7\%&90\%&0\%&93\%\\
\midrule
HSCSFL&78\%&92\%&41\%&80\%&89\%&77\%&48\%&94\%&90\%&85\%\\
\midrule
\midrule
HSCSFL(S3)&83\%&91\%&69\%&81\%&86\%&77\%&25\%&94\%&91\%&85\%\\
\midrule
\midrule
\multicolumn{11}{c}{\textbf{$FMNIST_{0.8}$}}\\
\midrule
Krum&69\%&90\%&41\%&86\%&78\%&0\%&48\%&91\%&80\%&92\%\\
\midrule
Median&73\%&91\%&47\%&85\%&80\%&34\%&43\%&95\%&71\%&0\%\\
\midrule
Tri Mean&74\%&90\%&48\%&85\%&79\%&20\%&35\%&92\%&88\%&94\%\\
\midrule
FL Trust&81\%&90\%&64\%&85\%&81\%&23\%&6\%&92\%&83\%&92\%\\
\midrule
HSCSFL&85\%&92\%&94\%&87\%&66\%&78\%&23\%&92\%&93\%&87\%\\
\midrule
\midrule
HSCSFL(S3)&86\%&92\%&53\%&80\%&93\%&78\%&17\%&93\%&93\%&87\%\\
\midrule
\bottomrule[1pt]
\end{tabular}
\label{tab6}
\end{center}
\end{table}

\subsection{Adaptive Experiments}
We consider the adversary could occupy 15\%, 25\%, and 35\% overall clients and further compare the existing FL methods and our proposed HSCSFL on FMNIST with a non-IID degree of 0.8 when attack strategy 2 is performed. Table \ref{CStab15} illustrates the testing accuracy of the attacked class under different FL methods and adversary occupations.

\begin{table}[h]
\caption{Illustration of the attacked class's accuracy under different FL methods against Label Flipping Attacks in FMNIST with non-IID degree as 0.8 when 15\%, 25\%, and 35\% clients are adversarial.}
\begin{center}
\begin{tabular}{c|c|c|c|c|c}
\toprule[1pt]
\midrule
\textbf{Adversary} & \textbf{Krum} & \textbf{Median} & \textbf{Tri Mean} & \textbf{FL Trust} & \textbf{HSCSFL}\\
\midrule
15\% & 53.7\% & 53.9\% & 28.9\% & 36.6\% & 71.1\%\\
\midrule
25\% & 0\% & 34.5\% & 0.8\% & 23.4\% & 78.0\%\\
\midrule
35\% & 0.8\% & 0\% & 4.1\% & 23.9\% & 74.6\%\\
\midrule
\bottomrule[1pt]
\end{tabular}
\label{CStab15}
\end{center}
\end{table}
The experimental results show that the existing FL methods drop the learning performance on the attacked class when increasing the occupation of Label Flipping Attack adversaries, while HSCSFL can effectively defend against the attack. Specifically, Krum and Trimmed Mean achieve around 0\% when the amount of adversary reaches 25\%. FL Trust shows a better defense performance when including more attackers while receiving around 36\% class testing accuracy under 15\% adversary and 24\% under 25\% and 35\% adversary.

\begin{figure}
\centering
{\includegraphics [width=0.32\textwidth] {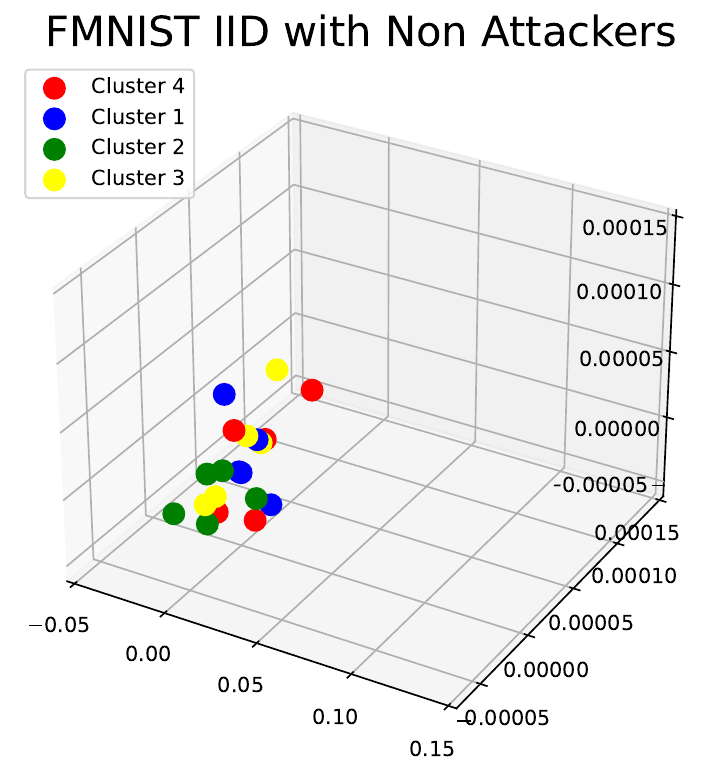}}
{\includegraphics [width=0.32\textwidth] {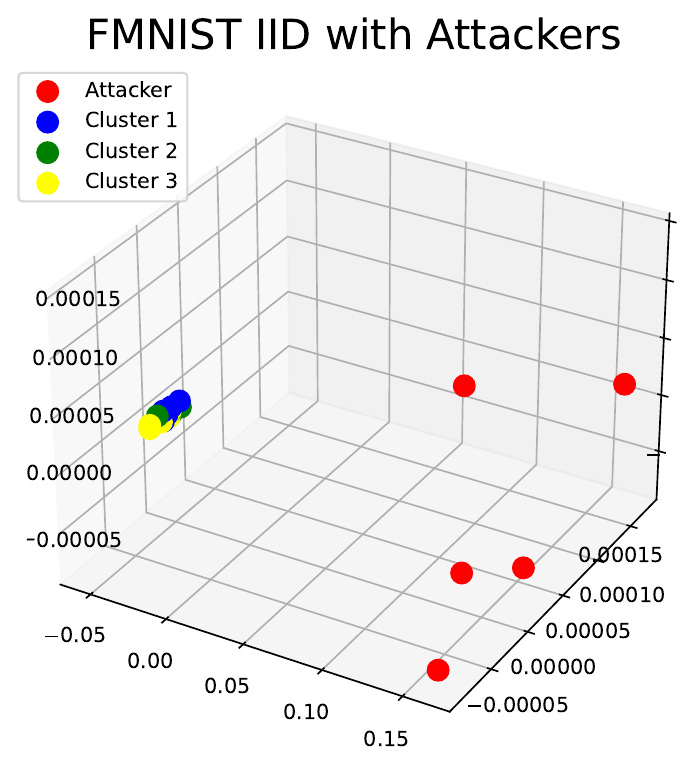}}
{\includegraphics [width=0.32\textwidth] {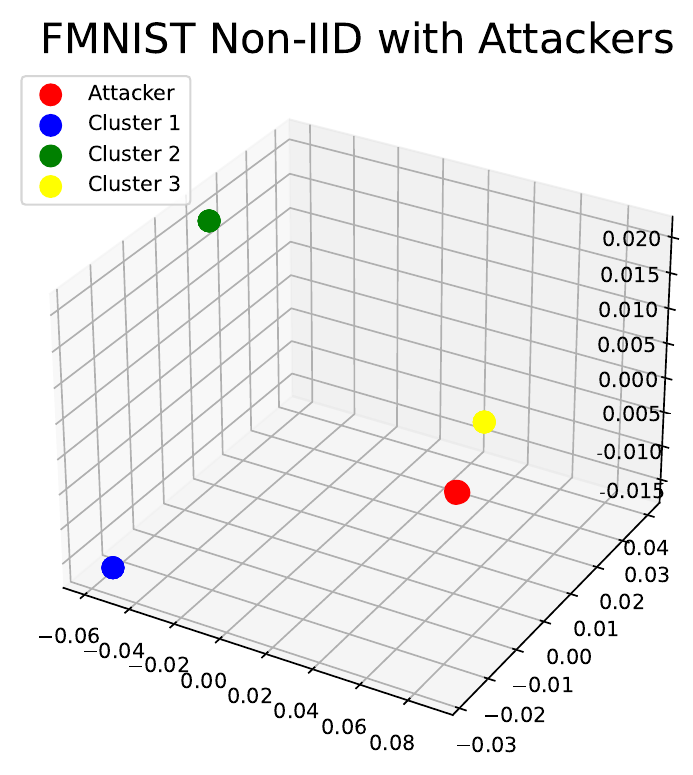}}
\caption{Illustration of the clustering outcome with Fashion MNIST in non-adversarial under IID setting, adversarial under IID setting, and adversarial under Non-IID setting.}
\label{figpca}
\end{figure}

\subsection{Effectiveness of clustering method}
We note that the existing work \cite{9622905} shows that clustering method could be introduced at the server to distinguish malicious gradients (under Label Flipping Attack) from benign clients in IID scenarios. However, whether the clustering method also shows effectiveness in Non-IID settings remains unclear. In this section, we investigate the effectiveness of the clustering method in IID with non-adversarial, IID with Label Flipping Attackers, and Non-IID with Label Flipping Attackers. We follow the scenario set in Section \ref{data}, use the $FMNIST_{0.9}$ as the a training data, attacking strategy 3 for performing attack and PCA (Principal Component Analysis) \cite{9622905} as the clustering method. To provide more information, we set the principal Component as 3.

Figure \ref{figpca} illustrates the clustering results for three different scenarios. In an IID setting with a non-adversarial scenario, participants (or gradients) are close due to the similar information carried by the training data. However, when Label Flipping Attacks are introduced in this IID setting, benign clients tend to maintain their close pattern. In contrast, the gradients of malicious clients deviate significantly, creating a noticeable distance from the benign gradients. This makes it feasible to effectively distinguish between them \cite{9622905}. However, due to the distinct information they carry, even benign clients exhibit varied behaviors and maintain significant distances from one another. This variation makes it challenging to distinguish malicious clients without additional information. 

\section{Discussion}
In this section, we provide the discussion around two key components for achieving HSCSFL robustness: 1) The Honest Score difference between the benign and malicious clients. 2) The evaluation dataset for generating the Honest Score.

\subsection{Honest Score Difference}
HSCSFL maintains the model robustness through the HSCS scheme, which assigns the suspicious clients lower scores and excludes them from the aggregation. Thus, ensuring the adversary achieves a lower honest score than the benign clients is the key to effectively performing HSCSFL. As the adversary generates the crafted gradient based on the same data distribution and volume in this paper, they generally receive the same honest score ($HS$) in each iteration. We monitor the honest score difference ($HS_d$) between benign clients and adversaries; specifically, the positive $HS_d$ indicates the attackers achieve lower $HS$ than the benign client and vice versa.

\begin{figure}[h]
\centering
{\includegraphics [width=0.49\textwidth] {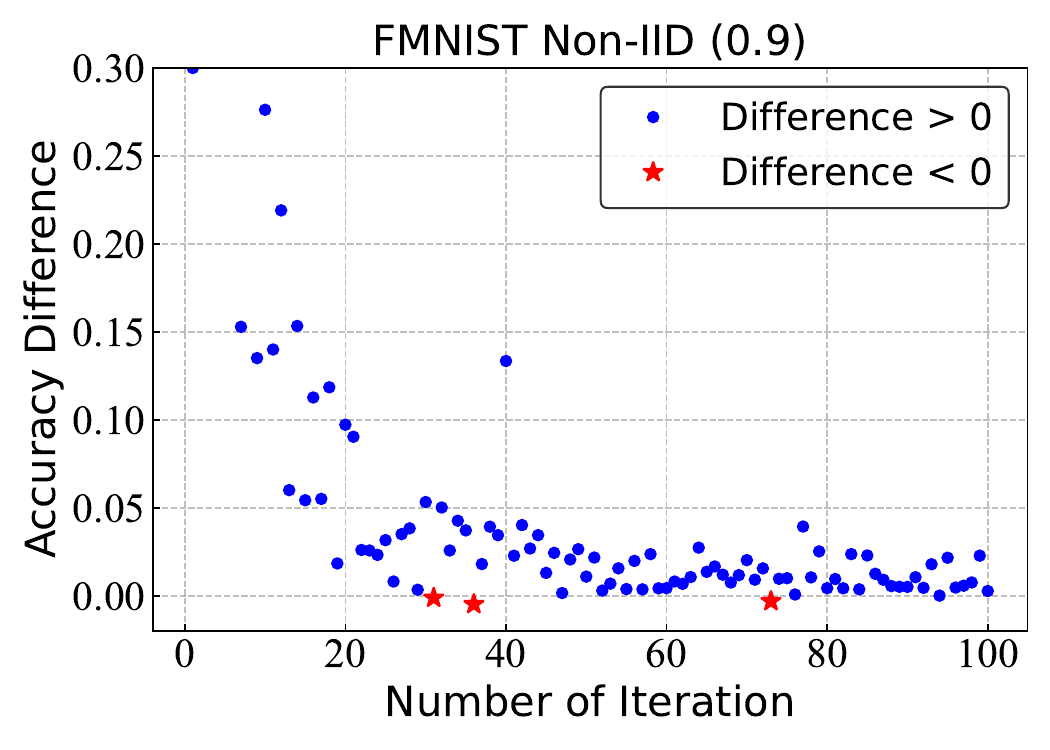}}
{\includegraphics [width=0.49\textwidth] {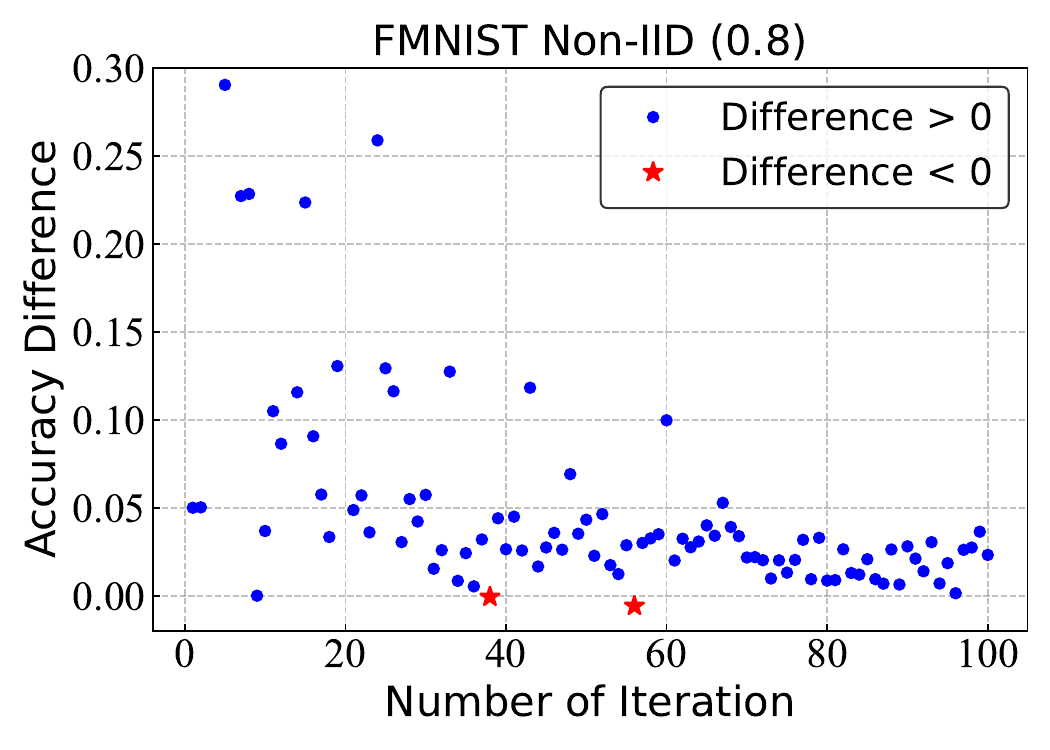}}

\caption{Illustration of the HS difference between adversary and benign clients in FMNIST (attack strategy 2) with different attack strategies. The blue point indicates the benign clients achieve a higher $HS$ and the difference is positive; the red star indicates the benign clients achieve a smaller $HS$ and the difference is negative.}
\label{fig8}
\end{figure}

Figure \ref{fig8} illustrates the $HS_d$ between benign clients and malicious clients in FMNIST with different non-IID degrees.  The results show that benign clients can generally keep a higher HS than the adversary during the learning process; in other words, the adversary can hardly participate in the aggregation and perform the attack. For instance, the adversary only participates in the aggregation two times under FMNIST when the non-IID degree is 0.8 and attack strategy 2 is performed. Recall the necessary condition for successfully performing a Label Flipping Attack includes the adversary should keep participating in the aggregation \cite{tolpegin2020data}, our HSCSFL can effectively defend against Label Flipping Attack and achieve a good learning performance.  We consider this HS gap to be “unavoidable” for performing the Label Flipping Attack. From the HSCS scheme perspective, the clients' gradients' performance loss in each class will be counted and collected as the $AccV$. As Our HSCSFL weights the class's accuracy $AccV$ dynamically based on the global model, the accuracy gap on the potential attacked class is scaled up, which results in the adversary achieving a lower $HS$ compared with the benign clients. Next, we provide a formal discussion of the organization of the “HS gap” and the effectiveness of the HSCS scheme.

Suppose the current learning task includes $C$ classes in total. The label flipping attackers select the class $m$ data to poison and generate the corresponding malicious gradients $g^{*}$, while the normal clients generate the benign gradients $g$. Suppose the adversary effectively attacks the global model and reduces the accuracy of class $m$ in iteration $t$, we have the following:

\begin{equation}
PerV^{t}(m) < PerV^{t}(c), c\in C, c\neq m \label{11}
\end{equation}
\begin{equation}
RisV = 1- PerV, RisV^{t}(m)>RisV^{t}(c), c\in C, c\neq m \label{12}
\end{equation}

To select the most honest clients to participate in $t+1$ round aggregation, the server calculates the HS for each client. The HS includes the HS for benign classes (c) and attacked class (m) which could be further divided by the following:
\begin{equation}
HS = HS(m) + HS(c), c\in C, c\neq m \label{13}
\end{equation}

Based on the equation \ref{6} and \ref{13}, the HS for the adversary and benign clients are shown as equation \ref{14} and \ref{15}, respectively.
\begin{equation}
HS^{t+1}_{g^*} = AccV^{t+1}_{g^*}(m) \cdot RisV^t(m) + AccV^{t+1}_{g^*}(c) \cdot RisV^t(c)\label{14}
\end{equation}
\begin{equation}
HS^{t+1}_{g} = AccV^{t+1}_{g}(m) \cdot RisV^t(m) + AccV^{t+1}_{g}(c) \cdot RisV^t(c)\label{15}
\end{equation}

Recall we consider the adversary and benign clients holding similar capability and quality training data (see Section 4.2), we have the following:
\begin{equation}
AccV^{t+1}_{g^*}(c) \cdot RisV^t(c) \simeq AccV^{t+1}_{g}(c) \cdot RisV^t(c)\label{16}
\end{equation}
\begin{equation}
HS^{t+1}_{g} - HS^{t+1}_{g^*} \simeq RisV^t(m)\cdot(AccV^{t+1}_{g}(m)-AccV^{t+1}_{g^*}(m)) \label{17}
\end{equation}

Following formula \ref{17}, we can find the HS gap between the benign clients and the adversary is the accuracy difference of their model performed on the attacked class ($m$) and then large-scaled by $RisV(m)$. The “large” we mentioned here is compared with the $RisV$ of other unattacked classes ($RisV(c)$), which is derived by \ref{12}. As long as the adversary performs label flipping attacks, its model unavoidably achieves lower accuracy on the class $m$ and is further large-scaled, resulting in a lower HS (we called HS gap) compared to benign clients, shown in Figure \ref{fig8}.

\subsection{Evaluation Dataset}
In HSCSFL we consider the evaluation dataset has been collected manually by the server (service provider) to guarantee its clean and unbiased. Here, we provide the discussion around the impact once the evaluation dataset has been polluted by an unknown attack or invasion.

We begin by formally expressing the Honest Score for both benign and malicious clients, where the evaluation dataset is biased or compromised. Following the formulas \ref{14} and \ref{15}, we reorganize the Honest Score as unaffected and affected to reflect the evaluation dataset of the corresponding classes are biased or polluted and clean, respectively.

\begin{equation}
HS = HS_{unaff} + HS_{aff}\label{18}
\end{equation}

Equation \ref{18} could be extended as \ref{19} and \ref{20} for benign and malicious clients, respectively. We use $c\in C-C^{*}$ to denote the class that are clean in the evaluation dataset, and use $c^{*} \in C^{*}$ to denote the polluted or biased class.

\begin{equation}
HS_{g}^{t+1} = \sum_{c=1}^{C-C^{*}}HS(c)_g^{t+1} + \sum_{c^{*}=1}^{C^{*}}HS(c^{*})_g^{t+1},c^{*} \neq c\label{19}
\end{equation}
\begin{equation}
HS_{g^{*}}^{t+1} = \sum_{c=1}^{C-C^{*}}HS(c)_{g^{*}}^{t+1} + \sum_{c^{*}=1}^{C^{*}}HS(c^{*})_{g^{*}}^{t+1},c^{*} \neq c\label{20}
\end{equation}
From equations \ref{19} and \ref{20} we can find that whether the attacked class $c$ belongs to clean classes ($C-C^{*}$) or biased/polluted classes ($C^{*}$) of the evaluation dataset will drive different outcomes, we further discusses these two situation separately.

\textbf{Situation I:} When attacked class $m \in C-C^{*}$, we have the following:
\begin{equation}
\sum_{c=1}^{C-C^{*}}HS(c)_g^{t+1} > \sum_{c=1}^{C-C^{*}}HS(c)_{g^{*}}^{t+1} \label{21}
\end{equation} 
On the other hand, both benign and malicious clients receive the similar magnitude carried by the bias of class $c^{*}$. Thus, we have the following:
\begin{equation}
\sum_{c^{*}=1}^{C^{*}}HS(c^{*})_g^{t+1}\simeq\sum_{c^{*}=1}^{C^{*}}HS(c^{*})_{g^{*}}^{t+1} \label{22}
\end{equation} 
Through combining \ref{21} and \ref{22}, we have the following:
\begin{equation}
\begin{aligned}
\sum_{c=1}^{C-C^{*}}HS(c)_g^{t+1}+\sum_{c^{*}=1}^{C^{*}}HS(c^{*})_g^{t+1} &> \sum_{c=1}^{C-C^{*}}HS(c)_{g^{*}}^{t+1}+\sum_{c^{*}=1}^{C^{*}}HS(c^{*})_{g^{*}}^{t+1}\\ 
HS_{g}^{t+1}&>HS_{g^{*}}^{t+1}\label{23}
\end{aligned}
\end{equation}
Consequently, even when the evaluation dataset is biased or polluted, our HSCSFL algorithm maintains its robustness and effectively filters out the malicious clients, provided that the attacked classes fall within the clean/unbiased classes in the evaluation dataset.

\textbf{Situation II:} When attacked class $m \in C^{*}$, the inequality \ref{21} will represented as the following:
\begin{equation}
\sum_{c=1}^{C-C^{*}}HS(c)_g^{t+1} \simeq \sum_{c=1}^{C-C^{*}}HS(c)_{g^{*}}^{t+1} \label{24}
\end{equation} 
However, since the evaluation data for the attacked class $m$ has been polluted or biased, false negatives and false positives will adversely affect the corresponding score. As a result, the relationship between $\sum_{c^{*}=1}^{C^{*}}HS(c^{*})_g^{t+1}$ and $\sum_{c^{*}=1}^{C^{*}}HS(c^{*})_{g^{*}}^{t+1}$ becomes uncertain, potentially leading to situations where malicious clients achieve a higher Honest Score ($HS$). 

In summary, while HSCSFL’s robustness may only be compromised in scenarios where the evaluation data for the attacked class $m$ is polluted or biased, we consider that such intrusive tampering with the evaluation is impractical in real-world scenarios. This is due to the evaluation dataset being meticulously prepared and securely maintained by the server or service provider, which is honest and plays the defender role.

\section{Conclusion}
In this paper, we evaluate the robustness of the existing FL methods when defending against “label flipping attack” in both IID and Non-IID scenarios. We find that although these FL methods maintain robustness in IID settings, they show a degrading performance and can not effectively aggregate different class information in Non-IID scenarios. 

To defend against the “label flipping attack” and achieve Byzantine robustness, we propose the HSCS scheme which enables the server to select the most honest clients and excludes the suspicious from the aggregation. We further provide the corresponding HSCSFL, and our evaluation shows that the HSCSFL can effectively select benign clients, aggregate the gradients' information from benign clients, defend against the “label flipping attack” in different Non-IID scenarios, and finally enhance the robustness of FL framework. We note that the current HSCSFL framework relies on knowledge about the number of adversaries. Therefore, enhancing the framework to accommodate an "unknown" number of attackers presents an intriguing direction for future research.

\bibliographystyle{unsrt}  
\bibliography{references}

\end{document}